\documentclass[12pt]{iopart}

\usepackage{iopams}  
\usepackage{graphicx}
\usepackage{iopams}

\expandafter\let\csname equation*\endcsname\relax

\expandafter\let\csname endequation*\endcsname\relax

\usepackage{amsmath}

\def\ER{Erd\H{o}s-R\'enyi }

\begin{document}

\title{Degree dependent transmission rates in epidemic processes}

\author{G. J. Baxter}
\ead{gjbaxter@ua.pt}
\author{G. Tim\'ar}
\address{Department of Physics \& I3N, University of Aveiro, Portugal}

\vspace{10pt}
\begin{indented}
\item[]\today
\end{indented}

\begin{abstract}
The outcome of SIR epidemics with heterogeneous infective lifetimes, or heterogeneous susceptibilities, can be mapped onto a directed percolation process on the underlying contact network. In this paper we study SIR models where heterogeneity is a result of the degree dependence of disease transmission
rates. 
We develop numerical methods to determine the epidemic threshold, the epidemic probability and epidemic size close to the threshold for configuration model contact networks with arbitrary degree distribution and an arbitrary matrix of transmission rates (dependent on transmitting and receiving node degree). For the special case of separable transmission rates we obtain analytical expressions for these quantities. 
We propose a categorization of spreading processes based on the ratio of the probability of an epidemic and the expected size of an epidemic, and demonstrate that this ratio  has a complex dependence 
on the degree distribution and the degree-dependent transmission rates. 
For scale-free contact networks and transmission rates that are power functions of transmitting and receiving node degrees, 
the epidemic threshold may be finite even when the degree distribution powerlaw exponent is below $\gamma < 3$.
We give an expression, in terms of the degree distribution and transmission rate exponents, for the limit at which the epidemic threshold vanishes.
We find that the expected epidemic size and the probability of an epidemic may grow nonlinearly above the epidemic threshold, with exponents that depend not only on the degree distribution powerlaw exponent, but on the parameters of the transmission rate degree dependence functions, in contrast to ordinary directed percolation and previously studied variations of the SIR model.
\end{abstract}

%
%
%
%
%


\section{Introduction}\label{intro}

Relatively simple compartmental models have proven to be remarkably effective representations of real disease spreading.
One of the most broadly studied epidemic models is the SIR model, due to its simplicity, applicability and tractability \cite{anderson1992infectious}.
In this model, agents exist in one of three states: Susceptible, Infected, or Recovered. At the beginning of the process, most agents are susceptible, while a few are infected. A susceptible individual, upon contact with an infected individual, may themselves become infected. Infected agents may, in turn, spontaneously recover, after which no further changes of state are possible. At the end of the process, all agents are either recovered (meaning they at some time contracted the infection) or susceptible (were never infected).
This model is useful for studying contagions to which immunity is usually acquired, so that repeated infection is negligible. The infection either dies out quickly, after infecting only a finite number of individuals, or, beyond a certain transmission threshold, an epidemic occurs, in which a finite fraction of the population is infected. One may study this phase transition by looking at the final state of the system.

A particularly informative generalisation has extended these models to a heterogeneous substrate -- i.e. the individuals are located on a heterogeneous contact network \cite{pastor-satoras2001epidemic,moreno2002epidemic, newman2002spread,keeling2005networks}. Studies of epidemic models on networks have examined the effect of degree distributions and other network structure \cite{pastor-satoras2001epidemic,moreno2002epidemic,pastor2015epidemic} and neighbor degree correlations \cite{boguna2002epidemic} and numerous aspects and generalisations \cite{pastor2015epidemic}. 
Most of these studies 
assume that, in the limit of a large network, infection begins from a vanishingly small but not finite
 number of sites, so that there are no fluctuations with regard to the initial growth of the infection. This allows the problem to be mapped to undirected bond-percolation on the same network \cite{grassberger1983critical}.
One result of this assumption is that heterogeneity in transmission rates does not affect the epidemic threshold or the size of the epidemic, so one may treat the problem assuming transmission rates are homogeneous \cite{newman2002spread}.

A few authors, however, have studied epidemics originating with a single infected site \cite{kenah2007network,miller2007epidemic,rogers2015assessing}.
In this case, one must consider not only the epidemic size but also the probability that it occurs. The bond percolation mapping is not sufficient, and instead a generalised directed percolation method must be used \cite{miller2007epidemic,kenah2007network}. Futhermore, heterogeneity in transmission rates now affects the outcome. 
A similar analysis for specific finite networks, using the cavity method (belief propagation or message passing) was given in \cite{rogers2015assessing}.

The phenomenon of so called super-spreaders in virus transmission \cite{stein2011super} highlights the importance of considering heterogeneity of transmission rates. 
Meanwhile
in \cite{moslonka2012weighting}, the authors argue that, in highly heterogeneous sexual contact networks, it is unrealistic to assume that the transmission risk per partnership is equal. Rather, an individual with numerous contacts does not transmit the infection to each partner with the same probability as an individual with few contacts. 
Similarly, the available contact network information may be aggregated over time. If real contacts are distributed in time, this will also reduce the transmission rates, and we would again expect rates that decline with increasing degree.
Generally, then, one  should not expect infection transmission probabilities in an epidemic process to be uniform in a heterogeneous network.

Heterogeneity in transmission rates on networks has usually been modelled using weighted edges \cite{karsai2006nonequilibrium, britton2011weighted, ferreira2016collective}.
In \cite{karsai2006nonequilibrium} the authors considered networks with powerlaw degree distributions, and edge weights proportional to the product of the end degrees raised to a negative power, as proposed in \cite{giuraniuc2005trading} applied to an Ising model. 
Explicit dependence of the transmission rates on node degrees was considered in an SIS model in \cite{joo2004behavior}, specifically considering saturation effects for large degrees. A more general degree dependence scheme was proposed in \cite{olinky2004unexpected}, where the important observation was made that such degree dependence can counteract the expected effects of scalefree networks, in particular that an infection threshold may persist even for degree distribution powerlaw exponents below three. 
This observation was reinforced for the SIR model in \cite{chu2011epidemic}, where a transmission rate proportional to the infected node degree and the susceptible degree, raised to different powers. 
Note that a $1/k$ dependence on degree corresponds to the contact process on networks \cite{castellano2006non}, and it has been shown that this dependence has the tendency to neutralise the effect of degree-heterogeneity on spreading processes \cite{olinky2004unexpected}.
Degree dependent transmission rates have also been considered in terms of a related rumour spreading model \cite{singh2013nonlinear}.

Here we consider the effects of arbitrary degree dependent transmission rates in a networked SIR model.
We consider transmission rates proportional to an arbitrary function $f(k,k')$ of the infective node degree $k$ and susceptible node degree $k'$.
%
The transmission probability may be different in different directions along the same edge.
We show that as a results the probability of an epidemic, and the expected size of the eventual epidemic (whose product then gives the expected outbreak size) may vary considerably. We classify processes as aggregation, dissemination or balanced processes according to the relative size of these quantities, and enumerate conditions under which each occurs.

We show that broad degree distributions (so called scalefree networks) have a strong effect on the epidemic threshold and the growth exponents of the epidemic probability and size above the threshold. These exponents 
depend both on the degree distribution and the functional form of the transmission rate degree dependence, illustrating a complex interaction between the network structure and the heterogeneous transmission rates.
In particular a finite threshold may persist even in networks with degree distribution decay exponent $\gamma$ less than three.
For the example function $f(k,k') = k^{\alpha}k'^{\beta}$ we give a condition for $\alpha$ and $\beta$ for which the network effect in scalefree networks is effectively removed, and obtain a condition for the epidemic threshold to be finite.


\section{Heterogeneous epidemic process}\label{definition}

We consider a generalization of the SIR compartmental epidemic model in which individuals are located on a heterogeneous undirected contact network. 
All individuals begin in the susceptible state ($S=1$). At time $t=0$ one individual, $i$, selected uniformly at random, is infected with a disease, changing to the infective state. After a time $\tau_i$ the agent moves to the recovered state, and may no longer spread or catch the disease. During the time that $i$ is infective
the disease is transmitted to a susceptible neighbor $j$ at a rate $\beta_{ij}$. 
Infected individuals may in turn pass the infection to further susceptible individuals to whom they are connected, before themselves recovering.
The transmission rate $\beta_{ij}$ from infective agent $i$ to a susceptible neighbor $j$ may be different for each $i$ and $j$. 
These transmission rates need not be symmetric, i.e., $\beta_{ij}$ does not necessarily equal $\beta_{ji}$.
Furthermore, the time $\tau_i$ during which a node $i$ remains infective may also be heterogeneous. For example, if we impose a constant recovery rate $\gamma$ for all agents, then $\tau_i$ is drawn from an exponential distribution.

The infection may die out after a small number of infections, or spread to a large fraction of the population.
In the final state, all agents are either susceptible or recovered, with the total fraction of recovered agents $R$ corresponding to the total size of the outbreak.
In the large population limit, in the homogeneous case $\beta_{ij} = \beta\,\, \forall \, i, j$ there is a well defined threshold $\beta = \beta_c$ beyond which the mean outbreak size is a finite fraction of the population, and we say that an epidemic has occurred. 
Generally, one wishes to identify this epidemic threshold, as well as the expected size of the resulting epidemic, and the probability that it occurs.

\label{analysis}

\subsection{Directed network mapping}\label{directed}

\begin{figure}[htpb!]
\centering
\includegraphics[width=0.6\columnwidth,angle=0.]{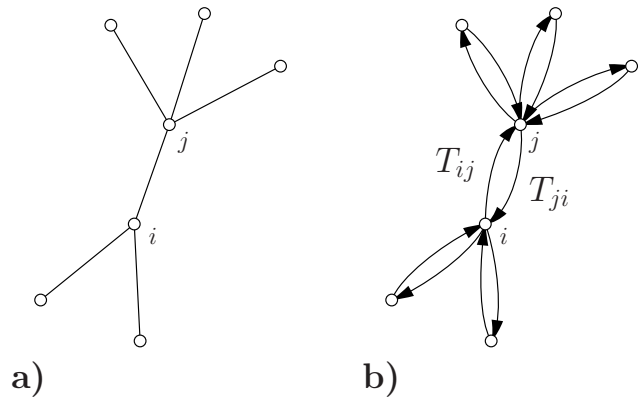}
\caption{a) A subgraph of an undirected network where a spreading process is to take place. b) Replacing each undirected link with a pair of directed links going in opposite directions. Each directed link is activated with probability $T_{ij}$
to give a correspondence between the spreading process and directed percolation.}
\label{fig:10}
\end{figure}

The probability that susceptible neighbor $j$ of infected site $i$ becomes infective before $i$ recovers is
\begin{align}
T_{ij} = 1 - e^{-\beta_{ij}\tau_i}.
\end{align}
Given that the transmission rates $\beta_{ij}$ are independent of the time at which $i$ becomes infected, or the state of the system, the possible final outcomes and their probabilities are determined only by the transmission probabilities $T_{ij}$ (see Figure \ref{fig:10}).

Let us consider the weighted directed network in which the directed edge from node $i$ to node $j$ is occupied with probability $T_{ij}$, and the directed edge from node $j$ to node $i$ is occupied with probability $T_{ji}$, with the probabilities in each direction not being necessarily equal. All nodes reachable following directed edges from the originally infected node will be infected in the outbreak, at one time or another, so the final size of the outbreak corresponds exactly to the size of the directed out-component connected to the originating node.
Similarly, the probability that a given node is ever infected corresponds to the probability that there is at least one directed path leading from the originating node to this node. This is equal to the size of the in-component connected to the node.
Thus the final state of the epidemic process corresponds to a directed percolation problem.

\begin{figure}[htpb!]
\centering
\includegraphics[width=0.6\columnwidth,angle=0.]{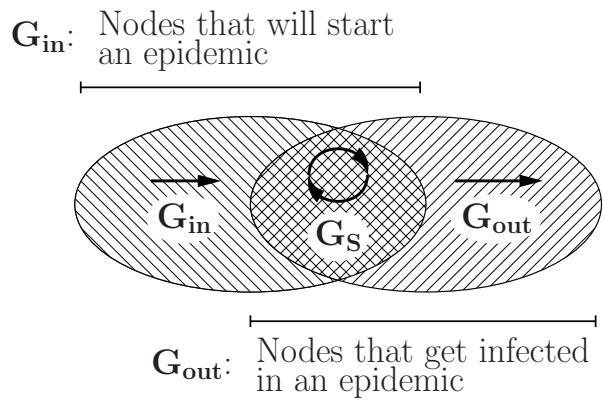}
\caption{The spreading process viewed as directed percolation. Nodes belonging to the giant in-component ($G_{in}$) will start an epidemic (if infected), and nodes belonging to the giant out-component ($G_{out}$) will get infected in an epidemic. Note that $G_S = G_{in} \cap G_{out}$ is the set of nodes that can both start an epidemic and get infected in one.}
\label{fig:20}
\end{figure}

The epidemic threshold, the appearance of a non-zero probability of an epidemic occurring, corresponds to the (simultaneous) appearance of a giant in- and giant out- component in the directed network.
All nodes in the giant out-component become infected if the infection started from a node in the giant in-component (see Figure \ref{fig:20}). The probability of the epidemic is therefore the probability that a randomly selected node belongs to the giant in-component, equal to the fraction of nodes $S_{in}$ it contains. Meanwhile, the expected size of the epidemic (given that it occurs) is given by the relative size of the giant out-component $S_{out}$.
Note that if all the transmission probabilities are symmetric (as in the standard SIR model, for example), then the problem essentially reduces to ordinary percolation.


We consider the case that the transmission rates $\beta_{ij}$ are a function of the degrees of $i$ and $j$:
\begin{align}
\beta_{ij} = \lambda f(k_i, k_j)
\end{align}
where $k_i$ is the number of neighbors of $i$, $\lambda$ is a tunable control parameter, and for simplicity we assume the mean value of $f(k,k')$ (over all links in the network) to be one.
Thus the probability that the infection ever passes from an infected node of degree $k$ and lifetime $\tau$ to a susceptible node of degree $k'$ is
\begin{align}\label{Tkk}
T_{k,k'}(\tau,\lambda) = 1 - e^{-\lambda f(k,k')\tau}.
\end{align}
Note that these probabilities are asymmetric, $T_{k,k'}$ is not necessarily equal to $T_{k',k}$.

We further assume that the infective lifetimes $\tau$ are iid random variables selected from a continuous distribution $R(\tau)$, whose mean we assume, without loss of generality, to be one.

\subsection{Epidemic probability and critical point}\label{INeqs}

Let us examine the size of an outbreak generated by the infection of a single individual.
We consider large sparse random networks drawn from the ensemble defined by the degree distribution $P(k)$ in the large size limit, commonly referred to as the configuration model. Such a network is asymptotically 
locally tree-like.
We define $x_k(\tau)$ to be the probability that only a finite number of individuals become infected via an edge emanating from a randomly selected infected node of degree $k$ and with infective lifetime $\tau$.
Note that therefore all edges emanating from the same node will be ascribed the same probability $x_k(\tau)$. 
Using the locally tree-like property of the network, we can write a recursive equation for $x_k(\tau)$ in terms of the equivalent probability for the child edges (edges not including the edge along which a node is reached) emanating from nodes neighboring the original node:
\begin{equation}\label{x_k}
x_k(\tau) = \int _0^{\infty} \! \! \! \! \! \! d\tau' R(\tau') \!
\sum_{k'=1}^{\infty} \! \! \frac{k'P(k')}{\langle k\rangle} \!\!
\left\{
1 \!- \!T_{k,k'}
(\tau)
[1-x_{k'}
(\tau')
^{k'-1}]
\right\}.
\end{equation}
The probability that a single infected node selected uniformly at random gives rise to an epidemic (outbreak infecting a finite fraction of the network) is then
\begin{align}\label{Sin}
S_{in}(\lambda) &= \sum_k P(k)\int _0^{\infty} \! \! \! \! \! \! d\tau R(\tau)
\left[1 - x_k
(\tau)
^{k}\right]\,.
\end{align}
This corresponds to the relative size of the giant in-component in the directed percolation mapping.

We can eliminate the dependence on $\tau$ by defining the probability  
$a_k(\lambda)$ that the number of individuals which become infected
via $k-1$ edges emanating from a randomly selected infected node of degree $k$ is {\em not} finite,
\begin{align}\label{G_k}
a_k(\lambda) &\equiv 
\int _0^{\infty} \! \! \! \! \! \! d\tau R(\tau)  [1-
x_k(\tau)^{k-1}]
\nonumber \\
&= 1 -  \int _0^{\infty} \! \! \! \! \! \! d\tau R(\tau) \left\{
    \sum_{k'=1}^{\infty} \! \! \frac{k'P(k')}{\langle k\rangle} \left[1 - T_{k,k'}(\tau)a_{k'}(\lambda)\right]\!
    \right\}^{k-1}
\end{align}
where we have written $a_k$ explicitly as a function of the control variable $\lambda$.
%
If the initial infection never leads to an epidemic, then $a_k = 0$ ($\forall k$) and hence $S_{in} = 0$. Increasing the transmission probabilities by increasing the parameter $\lambda$, we reach a threshold value $\lambda_c$ beyond which $S_{in} > 0$ and an epidemic becomes possible. 

To examine the behavior near the critical threshold, we can expand equation (\ref{G_k}) in $a_k$:
\begin{equation}\label{a_k_powers}
a_k(\lambda) =
\int _0^{\infty} d\tau R(\tau) 
\sum_{l=1}^{k-1}
    (-1)^{l-1}
    \binom{k-1}{l}
    \left[
        \sum_{k'} 
        \phi_{k,k'}(\tau,\lambda)
        a_{k'}(\lambda) 
     \right]^l
\end{equation}
where 
\begin{align}
\phi_{k,k'}(\tau,\lambda) = \frac{k'P(k')}{\langle k\rangle}T_{k,k'}(\tau,\lambda)\,.
\end{align}
The epidemic threshold can then be obtained by keeping only the linear term in equation (\ref{a_k_powers}):
\begin{equation}\label{a_k_linear}
a_k(\lambda) \approx
     (k-1) \!
    \int _0^{\infty} d\tau R(\tau) 
    \sum_{k'}
      \phi_{k,k'}(\tau,\lambda)
    a_{k'}(\lambda) \,.
\end{equation}
We can write this in matrix form as
\begin{align}\label{a_k_matrix}
\bf{a}(\lambda) = \bf{M}(\lambda)\bf{a}(\lambda)
\end{align}
where the elements of the matrix $\bf{M}(\lambda)$ are
\begin{align}\label{matrixM}
M_{k,k'}(\lambda) = (k-1)\frac{k'P(k')}{\langle k\rangle}\int _0^{\infty} \! \! \! \! \! d\tau R(\tau) 
     T_{k,k'}(\tau,\lambda).
\end{align}
Equation (\ref{a_k_matrix}) is a right eigenvector equation for matrix $\mathbf{M}(\lambda)$, with eigenvalue $1$. By definition $\mathbf{M}(\lambda)$ is a non-negative matrix and all components of $\mathbf{a}(\lambda)$ are positive. Assuming that $\mathbf{M}(\lambda)$ is irreducible, the Perron-Frobenius theorem states that there is only one strictly positive right eigenvector, and this corresponds to the largest eigenvalue. We thus have the condition for the critical point of equation (\ref{x_k}): the largest eigenvalue of matrix $\mathbf{M}(\lambda)$ must be $\nu_1 = 1$.

In general, due to the nonlinearity of equation (\ref{Tkk}), and the necessity to solve equation (\ref{a_k_linear}) simultaneously for all $k$, one must resort to numerical means to obtain the critical point in the most general case. 
In Appendix \ref{MatrixMethod} we outline a simple numerical strategy for finding the critical point to arbitrary precision.
Under certain, not very limiting, conditions, the calculation may be greatly simplified, and closed form solutions obtained, as we detail in Sections \ref{onlyINorOUT} and \ref{separable}, below.

\subsection{Epidemic size}\label{OUTeqs}

Now we turn to the probability that a randomly selected node is infected in an outbreak, which equals the relative expected outbreak size. We define the probability 
$y_k$
that only a finite set of individuals could infect a random node of degree $k$ via an incoming edge,
that is, the probability that an incoming edge connected to a node leads from a finite in-component.
The probability that an individual selected uniformly at random becomes infected during an outbreak, which gives the size of the epidemic, can then be written
\begin{align}\label{Sout}
S_{out}(\lambda) = \sum_k P(k) \left[1 - y_k(\lambda)^{k}\right]
\end{align}
Where $y_k$ is given by
\begin{align}\label{y_k}
y_k = 
\sum_{k'=1}^{\infty} \! \! \frac{k'P(k')}{\langle k\rangle}
\left\{
1 -  \int _0^{\infty} \! \! \! \! \! \! d\tau' R(\tau') 
T_{k',k}(\tau')[1-y_{k'}^{k'-1}]\,
\right\}.
\end{align}

Note that, because an individual's probability of being infected is not related to its (subsequent) infective lifetime, $y_k$ does not depend on the distribution of lifetimes, but only on the mean value. 

Similarly to before, we may expand equation (\ref{y_k}) in $b_k = 1-y_k$:
\begin{align}\label{b_k_powers}
b_k(\lambda) =
\sum_{k'=1}^{\infty} \! \! \frac{k'P(k')}{\langle k\rangle}
\overline{T}_{k',k}(\lambda)
\sum_{l=1}^{k'-1}
    (-1)^{l-1}
    \binom{k'-1}{l}%
        b_{k'}(\lambda)^l
\end{align}
where $\overline{T}_{k',k}(\lambda) \equiv \int 
 d\tau' R(\tau') T_{k',k}(\tau',\lambda)$.

Keeping only linear order terms yields
\begin{align}\label{b_k_matrix}
\bf{b}(\lambda) = \bf{M'}(\lambda)\bf{b}(\lambda)
\end{align}
with
\begin{align}
M'_{k,k'}(\lambda) 
= (k'-1)\frac{k'P(k')}{\langle k\rangle}\int _0^{\infty} \! \! \! \! \! d\tau R(\tau) 
     T_{k',k}(\tau,\lambda).
\end{align}
Matrices $\mathbf{M}(\lambda)$ and $\mathbf{M'}(\lambda)$ have the same eigenvalue spectrum, therefore the critical points of Eqs. (\ref{x_k}) and (\ref{y_k}) coincide, as expected---the giant in- and out-components emerge at the same value of $\lambda$.

\section{Aggregation, Balanced, and Dissemination processes}\label{process_types}

The difference between Eqs. (\ref{Sin}) and (\ref{G_k}) and Eqs. (\ref{Sout}) and (\ref{y_k}) means that the giant in-component of the directed network may be of a very different size to the giant out-component. Their sizes and the balance between them is determined by the degree distribution $P(k)$, the tranmission rate function $f(k,k')$ and the distribution of infective lifetimes $R(\tau)$.
This observation holds for any spreading process on a network in which transmission probabilities may be asymmetric.

\begin{figure}[htpb!]
\centering
\includegraphics[width=0.6\columnwidth,angle=0.]{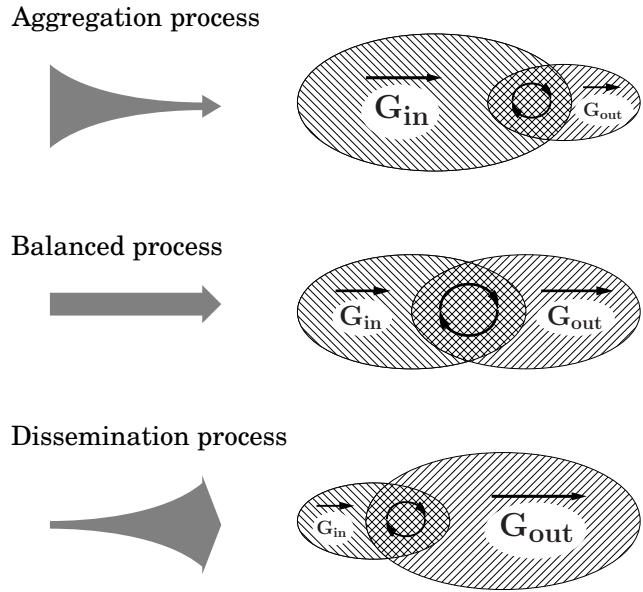}
\caption{Three different types of asymmetric spreading processes. The transmission probabilities $T_{k,k'}$ and the degree distribution $P(k)$ together determine what category a given process falls into. Symmetric processes (e.g., the SIR model) are perfectly balanced processes.}
\label{fig:30}
\end{figure}

One may classify a given process according to the relative sizes of the giant in- and out- components in the directed network mapping. If the giant in-component is larger than the giant out-component, we designate the process to be an {\em aggregation process}. The number of sites which, if infected, will lead to an epidemic is large (so the probability of an epidemic is large), even though the number of sites eventually infected (the epidemic size) is relatively small.
On the contrary, in a  {\em dissemination process}, in which the giant out-component is larger than the giant in-component, a relatively small number of possible infections give rise to an epidemic, but the epidemic may be large. When the two giant components are of similar size, we call this a {\em balanced process}. This classification is illustrated in Figure \ref{fig:30}

To quantify the relative sizes of the giant in- and out- components, we calculate the ratio of $S_{in}$ and $S_{out}$ with respect to $\lambda$ above the critical threshold
\begin{equation}
\Gamma = \frac{ S_{in}(\lambda_c) } { S_{out}(\lambda_c) }\,.
\end{equation}
The sizes $S_{in}(\lambda)$ and $S_{out}(\lambda)$ close to the critical point may be obtained by keeping second order terms in equation (\ref{a_k_powers}), again using a matrix formulation in terms of the principal eigenvectors of $\mathbf{M}$. We detail this procedure in Appendix \ref{GeneralSlopes}, resulting in Eqs. (\ref{slope_in_generic}) and (\ref{slope_out_generic}).
This calculation must be completed numerically in general, and is rather onerous. In the next sections we show that some simplifying assumptions allow us to reduce the problem to the solution of a single equation, allowing us to find closed form solutions.


\section{Dependence only on source or destination degree}\label{onlyINorOUT}

\subsection{Dependence only on source degree}
\label{onlyOUT}

We first consider the case that the infection rate depends only on the degree of the infected node, $f(k,k') = f(k)$. 
Thus 
\begin{equation}\label{Tk}
T_{k,k'}(\tau,\lambda) = 1 - e^{-\lambda \tau f(k)} \equiv T_{k}(\tau,\lambda)\,.
\end{equation}

Equation (\ref{G_k}) may be simplified by introducing the probability $A(\lambda)$ that transmission on a randomly selected edge leads to an epidemic (an infinite number of infected nodes),
\begin{equation}\label{A_defn}
A(\lambda) \equiv \sum_{k=1}^{\infty}\frac{kP(k)}{\langle k\rangle} a_k(\lambda).
\end{equation}
Substituting the righthand side of equation (\ref{G_k}) for $a_{k}$ we reduce the problem to a single equation:
\begin{align}
A(\lambda) = & 1 - \sum_{k=1}^{\infty}\frac{kP(k)}{\langle k\rangle} 
\int _0^{\infty} \! \! \! \! d\tau R(\tau)  
            \left[1 - T_k(\tau,\lambda)%
            A(\lambda) \right]^{k-1} \label{A}
\end{align}
With the probability of an epidemic originating from a single infection, equal to the size of the giant in-component, then given by
\begin{align}\label{Sin_onlyout}
S_{in} = 1 - \sum_{k=1}^{\infty}
P(k) \int _0^{\infty} \! \! \! \! d\tau R(\tau) 
    \left[1 - T_k(\tau,\lambda)%
    A(\lambda) \right]^{k}\,.
\end{align}

Expanding equation (\ref{A}) in $A$ and keeping only linear terms gives a relation for the critical point
\begin{align}\label{lamc_onlyout}
\left\langle k(k-1)\overline{T_k(\lambda_c)}
\right\rangle 
= \langle k\rangle\,.
\end{align}
where angled brackets indicate averages with respect to the degree distribution, while we use the overbar to indicate an average with respect to the lifetime distribution, 
\begin{align}\label{Tk_bar}
\overline{T_k(\lambda_c)} =  
1 - \int _0^{\infty} \!\!\!\! d\tau R(\tau)e^{-\lambda_c \tau f(k)}
\,.
\end{align}

Keeping also the second order term 
\begin{multline}
A(\lambda) \approx \sum_{k=1}^{\infty}\frac{k P(k)}{\langle k\rangle}\bigg\{
    \int _0^{\infty} \!\!\!\! d\tau R(\tau)
        (k-1)(1 - e^{-\lambda \tau f(k)}) A(\lambda)
        \\-
        \frac{1}{2}(k-1)(k-2)[(1 - e^{-\lambda \tau f(k)}) A(\lambda)]^2
        \bigg\}.\label{A_2nd}
\end{multline}
allows us to calculate the growth of $S_{in}$ above the critical point.
Writing averages over the degree distribution using angled brackets, and rearranging gives
\begin{equation}
     A(\lambda) = 
\frac{\langle k(k-1)\int \!\! d\tau R(\tau)(1 - e^{-\lambda \tau f(k)}) \rangle -
\langle k\rangle}
{\langle \frac{1}{2}k(k-1)(k-2)\int\!\! d\tau R(\tau)(1 - e^{-\lambda \tau f(k)})^2 \rangle}\,.\label{A_4th}
\end{equation}
Rewriting equation (\ref{Sin}) in terms of $A$, using (\ref{G_k}) we can write 
\begin{align}\label{Sin_A}
S_{in}(\lambda) &= \sum_k P(k) \int d\tau R(\tau) [1 - (1 - T_k(\tau)A)^k ]
 \approx \langle \overline{T_k(\lambda)}\rangle A
\end{align}
for small $A$. Thus, using (\ref{A_4th}), the growth of the giant in-component above the critical point is given by
\begin{align}\label{Sin_slope_k1}
S_{in}(\lambda) \approx
\frac{2
\langle k \overline{T(\lambda)}\rangle
[\langle k(k-1)\overline{T(\lambda)} \rangle -
\langle k\rangle]}
{\langle k(k-1)(k-2)
\overline{T^2(\lambda)}\rangle}\,.
\end{align}

To relate this result to the distance from the critical point, we write $\lambda = \lambda_c + \delta\lambda$, where $\lambda_c$ is the critical value of $\lambda$ found by solving equation (\ref{lamc_onlyout}). 
For small $\delta\lambda$, we can write
\begin{align}\label{exp_deltabeta}
e^{-\lambda\tau f(k)} \approx[1-\tau f(k)\delta\lambda]e^{-\lambda_c\tau f(k)} 
\end{align}
and substituting into equation (\ref{Sin_slope_k1}) gives
\begin{align}
S_{in}(\lambda) \approx
\delta\lambda
\frac{2
\langle k \overline{T_k(\lambda_c)}\rangle
[\langle k(k-1) f(k) \overline{\tau T_k(\lambda_c)} \rangle
}
{\langle k(k-1)(k-2)\overline{T_k(\lambda_c)^2}\rangle}\label{Sin_slope_k}
\end{align}
where as before a bar represents expectation value with respect to the infective lifetime distribution.

\medskip

Similarly, considering the giant out-component, substituting $T_k$ as defined in equation (\ref{Tk}) into equation (\ref{y_k}), we find that $y_k$ becomes independent of $k$, so we must solve only the single equation
\begin{align}\label{y}
y = 
\sum_{k'=1}^{\infty} \! \! \frac{k'P(k')}{\langle k\rangle}
\left[
1 - 
\overline{T_{k'}(\lambda)}(1-y^{k'-1})
\right]
\end{align}
with the size of the giant out-component then being simply
\begin{align}\label{Sout_y}
S_{out} = 1 - \sum_{k=1}^{\infty} P(k) y^k
\end{align}

Writing $b \equiv 1-y$, and expanding equation (\ref{y}) in $b$, we find
\begin{align}\label{b_powers}
b &= \sum_{k'=1}^{\infty} \! \! \frac{k'P(k')}{\langle k\rangle}
\overline{T_{k'}(\lambda)}[1-(1-b)^{k'-1}]\\
&= \sum_{k'=1}^{\infty} \! \! \frac{k'P(k')}{\langle k\rangle}
\overline{T_{k'}}
\sum_{l=1}^{k'-1}
(-1)^{l-1} \binom{k-1}{l} b^l\,.
\end{align}

Keeping only terms linear in $b$ returns the threshold condition, equation (\ref{lamc_onlyout}),
while keeping also quadratic terms gives the behavior of $b$ near the threshold:
\begin{align}
b &\cong \frac{
    \langle k(k-1) \overline{T_k(\lambda)}\rangle - \langle k \rangle
}
{
    \frac{1}{2}\langle  k(k-1)(k-2) \overline{T_k(\lambda)}\rangle
}\\
&=
\delta\lambda
\frac{
    2\langle k(k-1) f(k)\overline{\tau T_k(\lambda_c)}\rangle
}
{
    \langle  k(k-1)(k-2) \overline{T_k(\lambda_c)}\rangle
}
\end{align}
where we have again used Eqs. (\ref{exp_deltabeta}), and (\ref{lamc_onlyout}).

Returning to equation (\ref{Sout}) and linearising in $b$, we have
\begin{align}
S_{out}(\lambda) &= \sum_k P(k) \left[1 - (1-b)^{k}\right]
= \langle k\rangle b + \mathcal{O}(b^2)
\end{align}
thus
\begin{align}
S_{out}(\lambda) =
\delta\lambda
\frac{
    2\langle k\rangle\langle k(k-1) f(k)\overline{\tau T_k(\lambda_c)}\rangle
}
{
    \langle  k(k-1)(k-2) \overline{T_k(\lambda_c)}\rangle
}\,.\label{Sout_slope_k}
\end{align}

Combining the two results Eqs. (\ref{Sout_slope_k}) and (\ref{Sin_slope_k}), we find that the ratio of the sizes of the giant in-and out-components near the epidemic threshold is
\begin{align}\label{gamma_k}
\Gamma = \frac{
    \langle k \overline{T_k(\lambda_c)}\rangle \langle  k(k-1)(k-2) \overline{T_k(\lambda_c)}\rangle
}
{
    \langle k \rangle \langle  k(k-1)(k-2) \overline{T_k(\lambda_c)^2}\rangle
}\,.
\end{align}

\subsubsection{Results}\label{results_onlyOUT}

To explore the effect of degree dependent transmission probabilities on the spread of the epidemic, we consider a simple concrete form for the function $f(k,k')$
\begin{align}\label{fk_powers}
f(k,k') = k^{\alpha}
\end{align} 
where the exponent $\alpha$ may take positive or negative real values. When $\alpha=0$ we return to the original SIR model.
For simplicity, we will consider only the case of homogeneous infective lifetimes.
We may use the results of obtained above 
to calculate exact values for the sizes of the in-and out-giant components as a function of $\lambda$. 

\begin{figure}[htpb!]
\centering
\includegraphics[width=0.7\columnwidth]{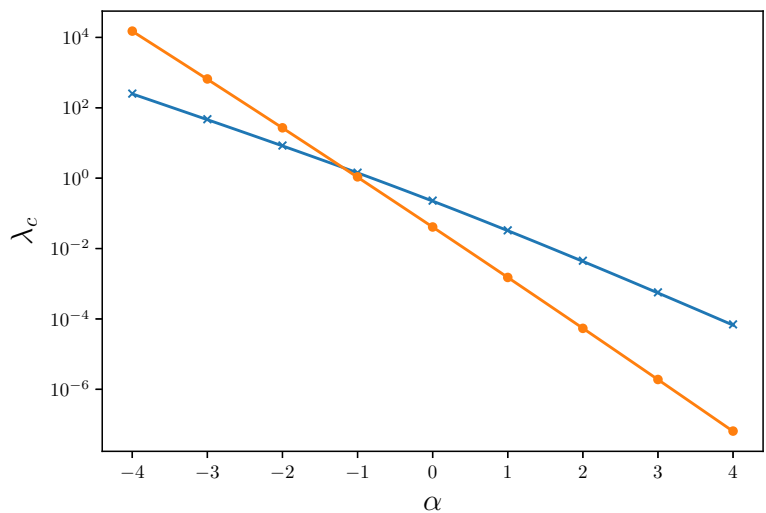}
\caption{Epidemic threshold as a function of $\alpha$ for \ER graphs, for transmission probabilities given by equation (\ref{Tk}). (blue, crosses) $\langle k\rangle = 5$, (orange, circles) $\langle k\rangle = 25$. Curves show theoretical solutions, markers show simulation results for networks with $N=10^6$ nodes.}
\label{fig_lamc_ER_onlyOUT}
\end{figure}

In Figure \ref{fig_lamc_ER_onlyOUT} we plot the epidemic threshold $\lambda_c$ as a function of $\alpha$. Theoretical results are in excellent agreement with simulations. We see that positive values of $\alpha$, that is, transmission rates positively correlated with infected degree, facilitate the emergence of an epidemic, reducing the threshold $\lambda_c$. Negative correlations, on the other hand, delay the transition. 
We see also that the values of $\lambda_c$ crossover close to $\alpha=-1$. Notice that at this value, the total infectivity of a node is constant, irrespective of its degree. Thus the effect of the degree distribution is largely counteracted.
From (\ref{lamc_onlyout}) and (\ref{Tk_bar}) one may show that if $\alpha \leq -1$ then $\lambda_c > 1$ for any degree distribution. To see this, consider that 
$1-e^{-k^{\alpha}} \leq 1-e^{-k^{-1}} < 1/(k-1) \forall k$. Thus the left hand side of (\ref{lamc_onlyout}) will be less than the right hand side if $\lambda_c=1$, requiring that $\lambda_c > 1$ for the equality to hold.
One may show using Jensen's inequality that this is true for any infective lifetime distribution.

\begin{figure}[htpb!]
\centering
\includegraphics[width=0.7\columnwidth]{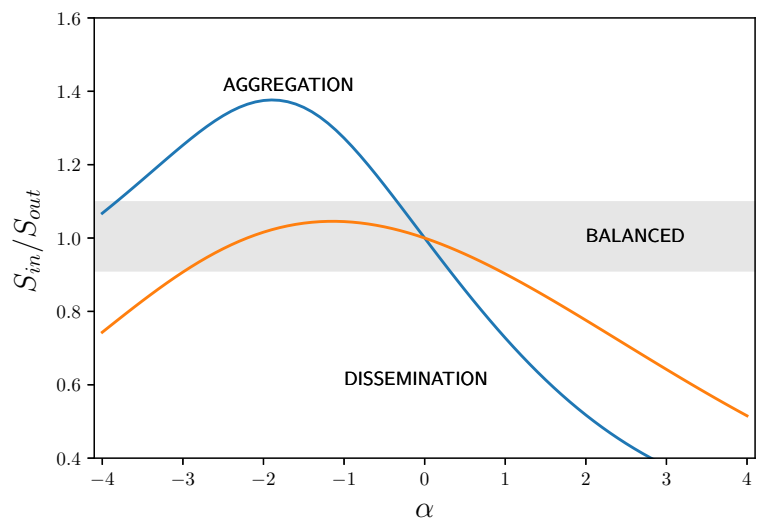}
\caption{Ratio of giant component sizes as a function of $\alpha$ for \ER graphs, for transmission probabilities given by equation (\ref{Tk}). (blue) upper at left $\langle k\rangle = 5$, (orange) lower at left $\langle k\rangle = 25$.}
\label{fig_ratio_ER_onlyOUT}
\end{figure}

The parameter $\alpha$ also has a dramatic effect on the relative sizes of $S_{out}$ (corresponding to the mean outbreak size, given that it occurs) and $S_{in}$ (the probability that an epidemic occurs), Figure \ref{fig_ratio_ER_onlyOUT}. For values of $\alpha$ close to zero, the process is balanced, as in the standard SIR model. For large positive $\alpha$, when transmission is enhanced for highly connected individuals, the process typically becomes disseminating, with a relatively small probability that a randomly located initial infection infects a large fraction of the population, but large infections when it does happen.
For negative $\alpha$, such as in our example of an infection for which highly connected individuals are less likely to infect any one particular neighbor, the process may become aggregating, however the trend is not monotonic, and the process may even be again disseminating.

\subsection{Dependence only on destination degree}\label{onlyIN}

In the case that the infection rate depends only on the degree of the node to be infected, $f(k,k') = f(k')$.

Substitution into equation (\ref{y_k}) gives
\begin{align}\label{y_konlyin}
y_k = 1 - 
 \overline{T_{k}}
\sum_{k'=1}^{\infty} \! \! \frac{k'P(k')}{\langle k\rangle}
 [1-y_{k'}^{k'-1}] 
\end{align}
and
letting $y_k = 1 - \overline{T_{k}}B(\lambda)$ we have
\begin{align}
B 
= 1 - \sum_{k'=1}^{\infty} \! \! \frac{k'P(k')}{\langle k\rangle}
 [1-\overline{T_{k}} B]^{k'-1}\,
\end{align}
which is identical to equation (\ref{A})  for $A$, in the case of dependence only on source degree when all lifetimes  are equal.
Then
\begin{align}\label{Soutonlyin}
S_{out}(\lambda) = \sum_k P(k) \left\{1 - [1 - \overline{T_{k}}B(\lambda)]^{k}\right\}
\end{align}
which matches equation (\ref{Sin_A}) for $S_{in}$ for only source dependence, again in the case that all lifetimes are equal.
Thus the behavior of $S_{out}$ near the threshold is the same as that given by equation (\ref{Sin_slope_k1}).

Making the same substitution $f(k,k') = f(k')$ in equation (\ref{G_k}) gives
\begin{align}\label{a_konlyin}
a_k(\lambda) 
&= 1 -  \!\int _0^{\infty} \! \! \! \! \! \! d\tau R(\tau) \left\{
    1 - \!\sum_{k'=1}^{\infty} \! \! \frac{k'P(k')}{\langle k\rangle} T_{k'}(\tau)a_{k'}(\lambda)
    \right\}^{k-1}.
\end{align}
Let $C(\tau) \equiv \sum_{k'=1}^{\infty} \! \! \frac{k'P(k')}{\langle k\rangle} T_{k'}(\tau)a_{k'}(\lambda)$,
then
\begin{align}\label{C_onlyin}
C(\tau)
&= 
\sum_{k'=1}^{\infty} \! \! \frac{k'P(k')}{\langle k\rangle} T_{k'}(\tau)
\left\{\!
1 \!- \! \! \int _0^{\infty} \! \! \! \! \! \! d\tau' R(\tau') [
    1\! -\! C(\tau')]^{k'-1} 
    \right\}.
\end{align}
Expanding in powers of $C$, and taking the average with respect to the lifetime distribution gives
\begin{align}
\overline{C}
&\approx
\sum_{k'=1}^{\infty} \! \! \frac{k'P(k')}{\langle k\rangle} \overline{T_{k'}}
[
     (k'-1)\overline{C} - \frac12(k'-1)(k'-2)\overline{C^2} + ...
    ] \label{Coverlineexpand}
\end{align}
taking the limit $\overline{C} \to 0$ and ignoring the quadratic term, we immediately recover the criterion for the critical point, which has the same formas in the case of dependence only on source degree:
\begin{align}
\langle k\rangle  =  { \langle k(k-1) \overline{T_{k}} \rangle}\,.
\end{align}

Because the $\tau$ dependent $T$ must remain within the summation over $k´$ in equation (\ref{C_onlyin}), it is not straightforward to obtain an expression for the growth of $S_{out}$ above the critical point.
However, in the case of homogeneous lifetimes, we may define and  defining $z =     1 - C$, which then obeys
\begin{align}
z(\lambda) 
&= 1 - \sum_{k'=1}^{\infty} \! \! \frac{k'P(k')}{\langle k\rangle} T_{k'} [1 -   z(\lambda)^{k-1}]
\end{align}
which is the same $k$ independent equation that $y$ obeys for only source dependence. Furthermore
\begin{align}\label{Sin}
S_{in}(\lambda) 
&= 1 - \sum_k P(k) z^k
\end{align}
which is identical in form to equation (\ref{Sout_y}).

Thus, when there is no lifetime dependence, the results for destination dependence are identical to those for source dependence under exchange of labels $in$ and $out$.
This is in fact clear from the symmetry of the process in this case.
 If we imagine reversing the direction of transmission, that is reversing all the directed edges in the directed network mapping, it is clear that the size of the giant out-component in the present case is exactly equal to the size of the giant in-component in the previous case. That is, we may obtain the desired results by exchanging the labels in- and out-.

Considering the function form $f(k,k') = k'^{\beta}$, the results illustrated in Figure \ref{fig_lamc_ER_onlyOUT} may be applied by simply exchanging $\beta$ for $\alpha$.
In other words, the effect in the threshold of an infector dividing his or her energy between neighbors ($\alpha < 0$) is equivalent to a susceptible individual dividing his or her contacts between neighbors ($\beta < 0$).
Similarly, the classification of the process as aggregation or dissemination, for example as in Figure \ref{fig_ratio_ER_onlyOUT} are inverted. 


\section{Dependence on both source and destination degrees}\label{separable}

In the completely general case, we cannot simplify the problem as we have done above, defining a new variable that is independent of degree and thus reducing the problem to the solution of a single equation for the giant in- and one for the size of the giant out-component. Instead, to obtain an exact solution, one must solve the full set of equations as defined in Section \ref{analysis}.
One may obtain, however, approximate solutions in simple closed forms, under certain simplifying assumptions.

Let us assume that $f(k,k')$ is separable, 
\begin{equation}
f(k,k') = g_k h_{k'}
\end{equation}
Further assuming that $\lambda f(k,k')$ is small, one may then linearise the exponential function in $T_{k,k'}$, allowing it too to be written in separable form:
\begin{equation}\label{Tkk_separated_f}
T_{k,k'}(\tau,\lambda) = 1 - e^{-\lambda \tau g_k h_{k'}} \approx \lambda \tau g_k h_{k'}
\end{equation}

Substituting into equation (\ref{a_k_linear}) 
and defining
\begin{equation}
A \equiv \frac{1}{\langle k h_k\rangle} \sum_{k'} P(k')k'h_{k'}(1-G_{k'}),
\label{A_sep_defn}
\end{equation}
which is independent of $k$, one must then only solve
\begin{align}\label{A_separable}
A = 1 -  \sum_{k=1}^{\infty}  \frac{P(k)kh_k}{\langle kh_k\rangle}
\int \!\! d\tau R(\tau)
\left[ 1 - \lambda \tau \frac{\langle kh_k\rangle}{\langle k\rangle}g_k A \right]^{k-1}
\end{align}
with the size of the giant in-component being then given by
\begin{align}\label{Sin_separable}
S_{in} = 
1 - \sum_{k=1}^{\infty} P(k) 
\int \!\! d\tau R(\tau)
\left[ 1 - \lambda \tau \frac{\langle kh_k\rangle}{\langle k\rangle}g_k A \right]^{k}\,.
\end{align}

Linearising equation (\ref{A_separable})
one immediately obtains an expression for the epidemic threshold
\begin{align}
\lambda_c = \frac{\langle k\rangle}{\langle k(k-1)g_kh_k\rangle}.
\label{lamc_separable}
\end{align}
Note that this is independent of the lifetime distribution $R(\tau)$ (we have assumed a mean lifetime $\overline{\tau} = 1$).

Similarly, beginning from equation (\ref{y_k}), we may define
\begin{align}\label{B_sep_defn}
B \equiv  1 - \frac{1}{\langle k g_k\rangle} \sum_{k'} P(k')k' g_{k'}y_{k'}^{k'-1}
\end{align}
which gives 
\begin{align}\label{B_separable}
B =  1 - \frac{1}{\langle k g_k\rangle} \sum_{k'} P(k')k' g_{k'}
\left[1 - \lambda h_{k'} \frac{\langle k g_k\rangle}{\langle k \rangle}B \right]^{k'-1}
\end{align}
and the size of the giant out-component may then be found via
\begin{align}\label{Sout_separable}
S_{out} = 1 - \sum_{k} P(k)
\left[1 - \lambda h_k \frac{\langle k g_k\rangle}{\langle k \rangle}B \right]^{k}\,.
\end{align}

To find the growth of the giant components with $\lambda$ above the critical threshold $\lambda_c$, we use the same method as before.
For simplicity we assume all individuals have $\tau=1$, although the more general case may be treated in the same way.
Truncating equation (\ref{A_separable}) to second order in $A$, writing $\lambda = \lambda_c + \delta\lambda$ and using the expression equation (\ref{lamc_separable}) for $\lambda_c$, we find
\begin{align}\label{A_growth_sep}
A \approx \delta\lambda \frac{2 \langle k(k-1) g_kh_k \rangle^3}{\langle k\rangle \langle k h_k\rangle \langle k(k-1)(k-2)g_k^2h_k \rangle}
\end{align}
and using equation (\ref{Sin_separable}) we obtain
\begin{align}\label{Sin_growth_sep}
S_{in} \approx \delta\lambda 
\frac{2 \langle k g_k\rangle}{\langle k\rangle}
\frac{2 \langle k(k-1) g_kh_k \rangle^2}{\langle k(k-1)(k-2)g_k^2h_k \rangle}\,.
\end{align}

Following a similar procedure with respect to the giant out-component gives
\begin{align}\label{B_growth_sep}
B \approx \delta\lambda \frac{2 \langle k(k-1) g_kh_k \rangle^3}
{\langle k\rangle \langle k g_k\rangle \langle k(k-1)(k-2)g_k h_k^2 \rangle}
\end{align}
and using equation (\ref{Sout_separable}) we obtain
\begin{align}\label{Sout_growth_sep}
S_{out} \approx \delta\lambda 
\frac{2 \langle k h_k\rangle}{\langle k\rangle}
\frac{2 \langle k(k-1) g_k h_k \rangle^2}{\langle k(k-1)(k-2)g_k h_k^2 \rangle}\,.
\end{align}

Thus the ratio of the sizes of the two giant components just above the epidemic threshold is
\begin{align}\label{ratio_separable}
\Gamma = \frac{S_{in}(\lambda)}{S_{out}(\lambda)} 
= \frac{\langle k g_k\rangle}{\langle k h_k \rangle}
 \frac{\langle k(k-1)(k-2) g_k h_k^2\rangle}{\langle k(k-1)(k-2) g_k^2 h_k \rangle}\,.
\end{align}


\subsection{Results}\label{results_BOTH}

To explore the effect of degree dependent transmission probabilities on the spread of the epidemic, we consider a simple concrete form for the function $f(k,k')$
\begin{align}\label{fkk_powers}
f(k,k') = k^{\alpha}k'^{\beta}
\end{align} 
where the exponents $\alpha$ and $\beta$ may take positive or negative real values. When $\alpha=\beta=0$ we return to the original SIR model. When one or the other is zero, we recover the results of the previous section.
We use the linear approximation to the transmission probability, equation (\ref{Tkk_separated_f}).
 For rapidly decaying degree distributions (i.e. non powerlaw tails) this is valid for large mean degree.

Note that the functions $g_k$ and $h_k$ appear together in the expression for $\lambda_c$, equation (\ref{lamc_separable}), that is, the prediction for the critical threshold depends only on the combination $\alpha+\beta$ and not on the two parameters separately.
As can be seen in Figure \ref{fig_lamc_separable}, the exact thresholds, obtained by the method described in Appendix \ref{MatrixMethod} indeed vary very little with $\beta$ (or $\alpha$) for a given value of $\alpha+\beta$ for large mean degree (\ER network with $\langle k\rangle = 25$ in this case). The approximation of equation (\ref{lamc_separable}) agrees very well with the exact results.
Even for relatively small mean degree, the approximation is good.
The reason for this is that, even expanding $T_{k,k'}$ beyong linear terms (see equation (\ref{Tkk_separated_f})), one finds that the next order solution for $\lambda_c$ also depends only on moments of $\alpha+\beta$.

\begin{figure}[htpb!]
\centering
\includegraphics[width=0.48\columnwidth]{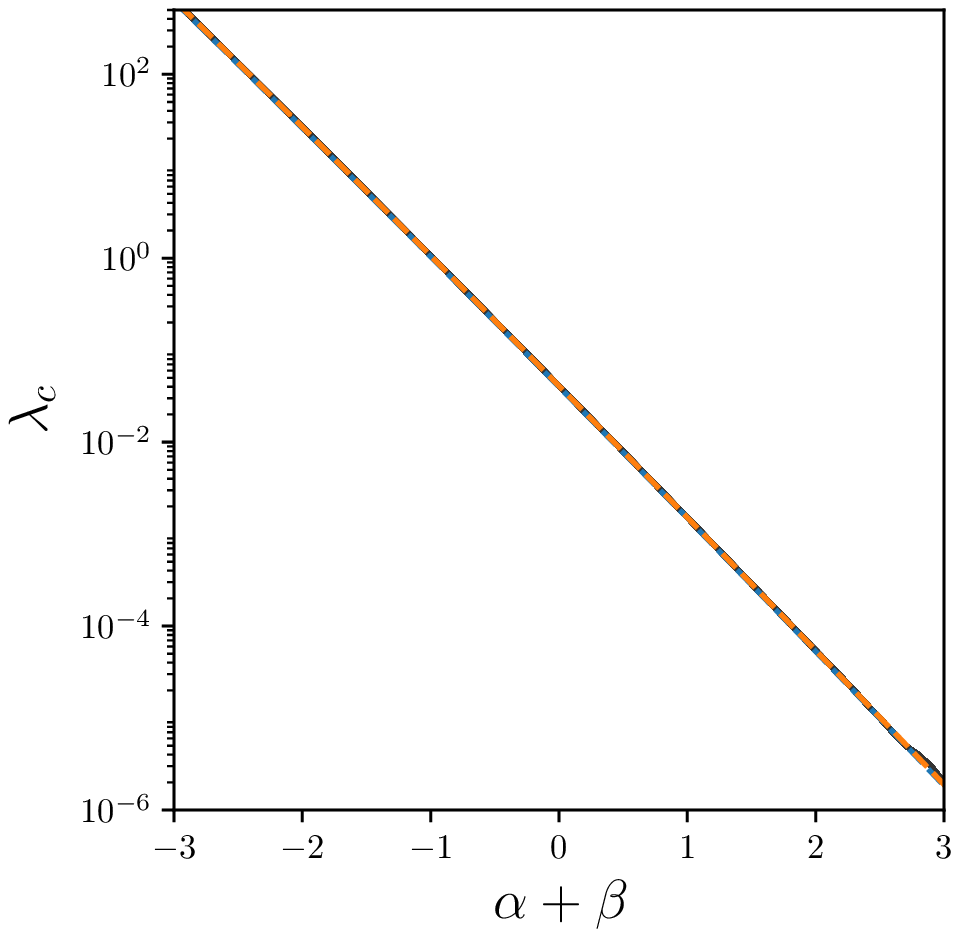}
\includegraphics[width=0.48\columnwidth]{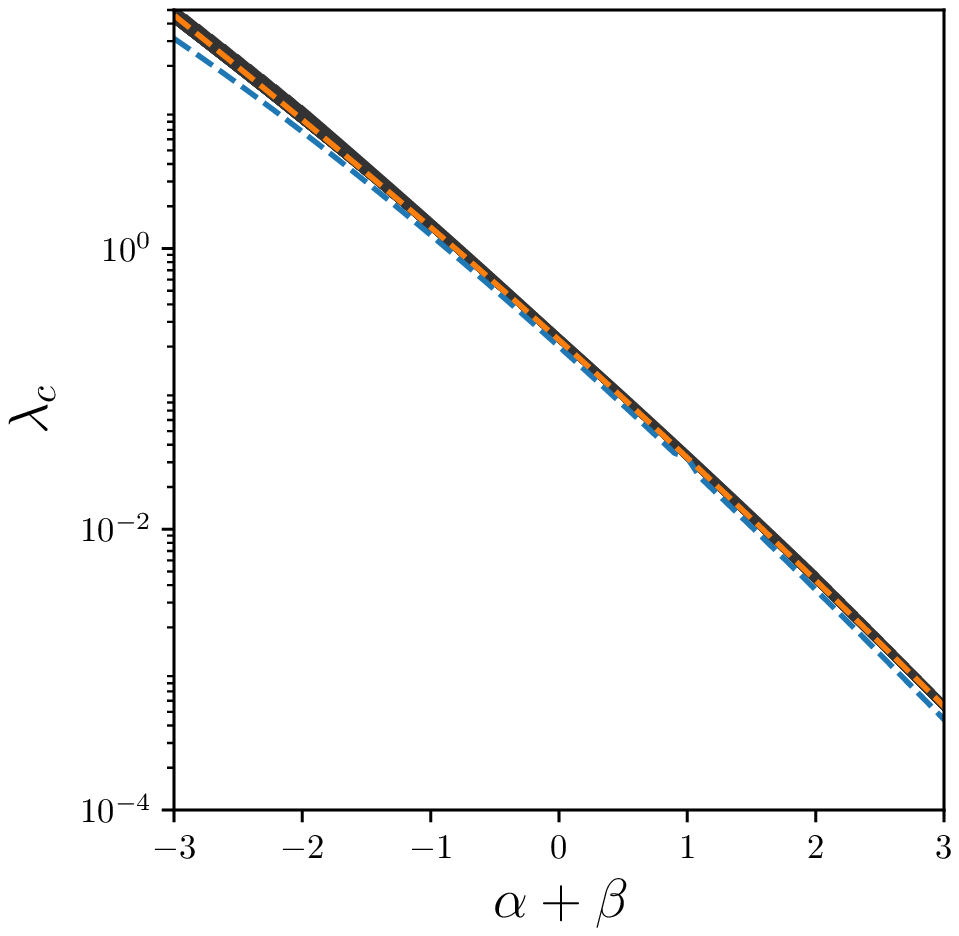}
\caption{Epidemic threshold $\lambda_c$ as a function of $\alpha + \beta$ in \ER graphs. Left panel: $\langle k\rangle = 25$, right panel $\langle k\rangle = 5$. 
Dark grey curves are matrix method (effectively exact) for various values of $\beta$ in the range $[-2,2]$. Blue dashed: linear approximation for $\beta= 1$. Orange curve: exact result for $\beta=0$.
}
\label{fig_lamc_separable}
\end{figure}

The initial growth of the giant components above the threshold, the rates at which the probability and expected size of the epidemic grow with $\lambda_c$, on the other hand, have a dependence on both $\alpha$ and $\beta$ which cannot be reduced to a single dimension, even in the leading order expansion, as can be seen in Eqs. (\ref{Sin_growth_sep}) and (\ref{Sout_growth_sep}) in which $g_k$ and $h_k$ appear not only as the product $g_kh_k$, but also as $g_kh_k^2$ or $g_k^2h_k$.

\begin{figure*}[htpb!]
\centering
\includegraphics[width=0.49\columnwidth]{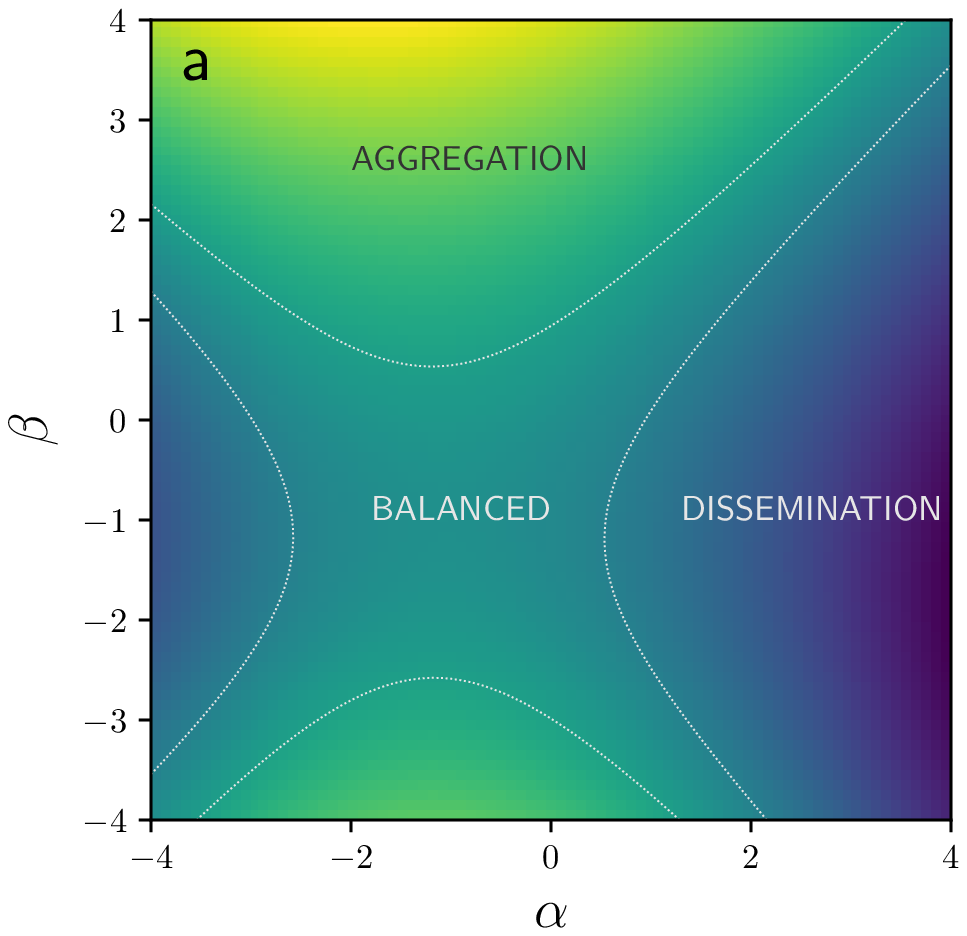}
\includegraphics[width=0.49\columnwidth]{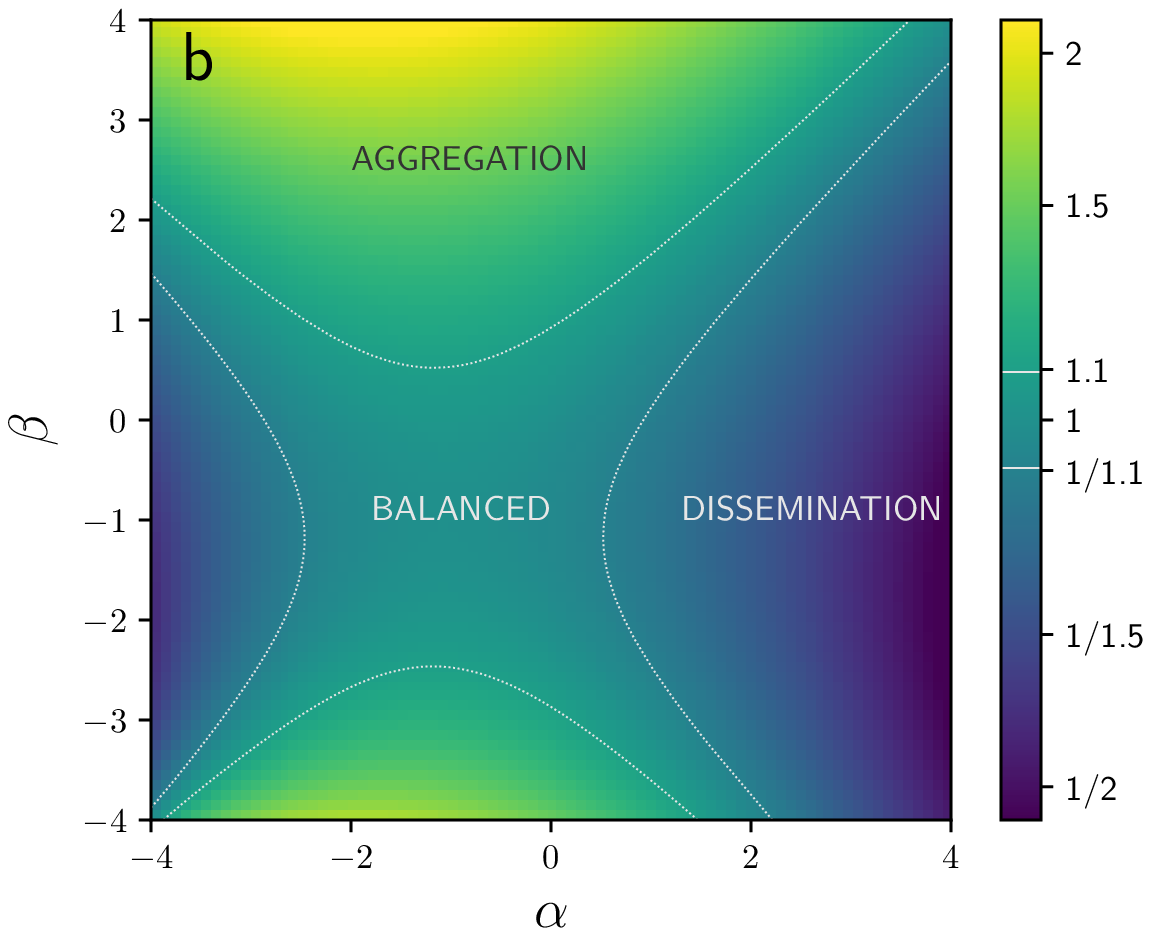}
\includegraphics[width=0.49\columnwidth]{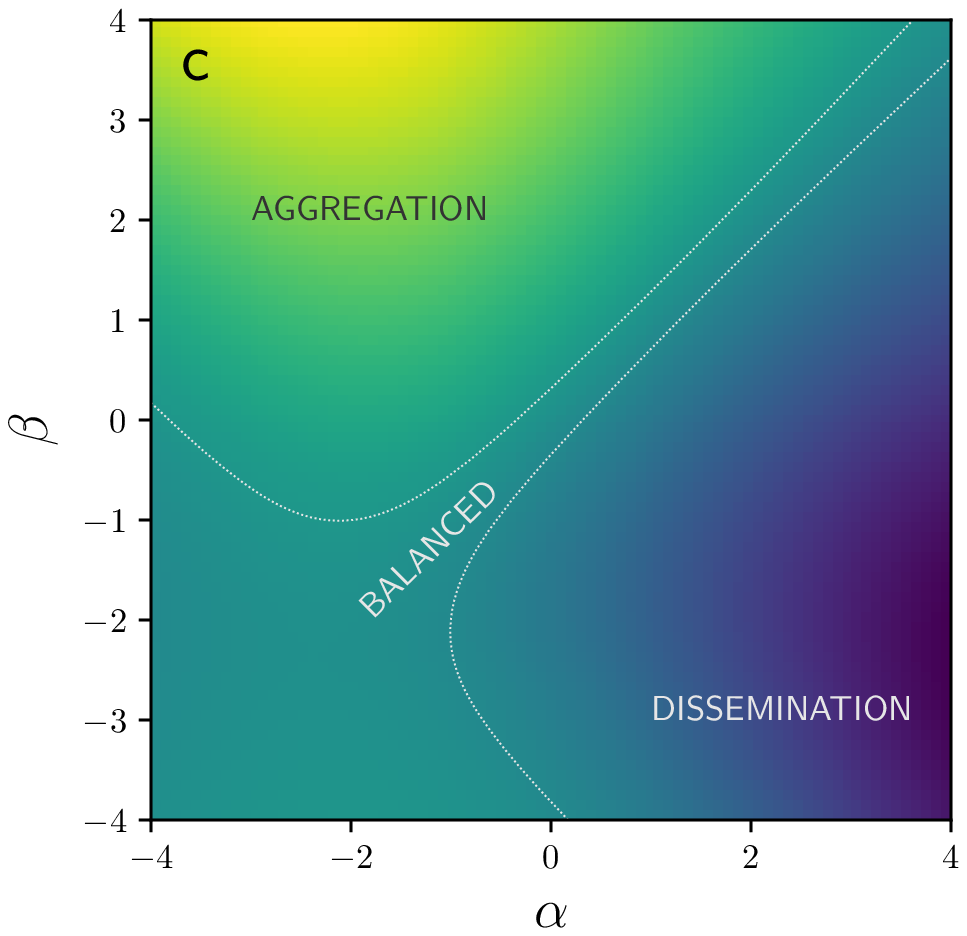}
\includegraphics[width=0.49\columnwidth]{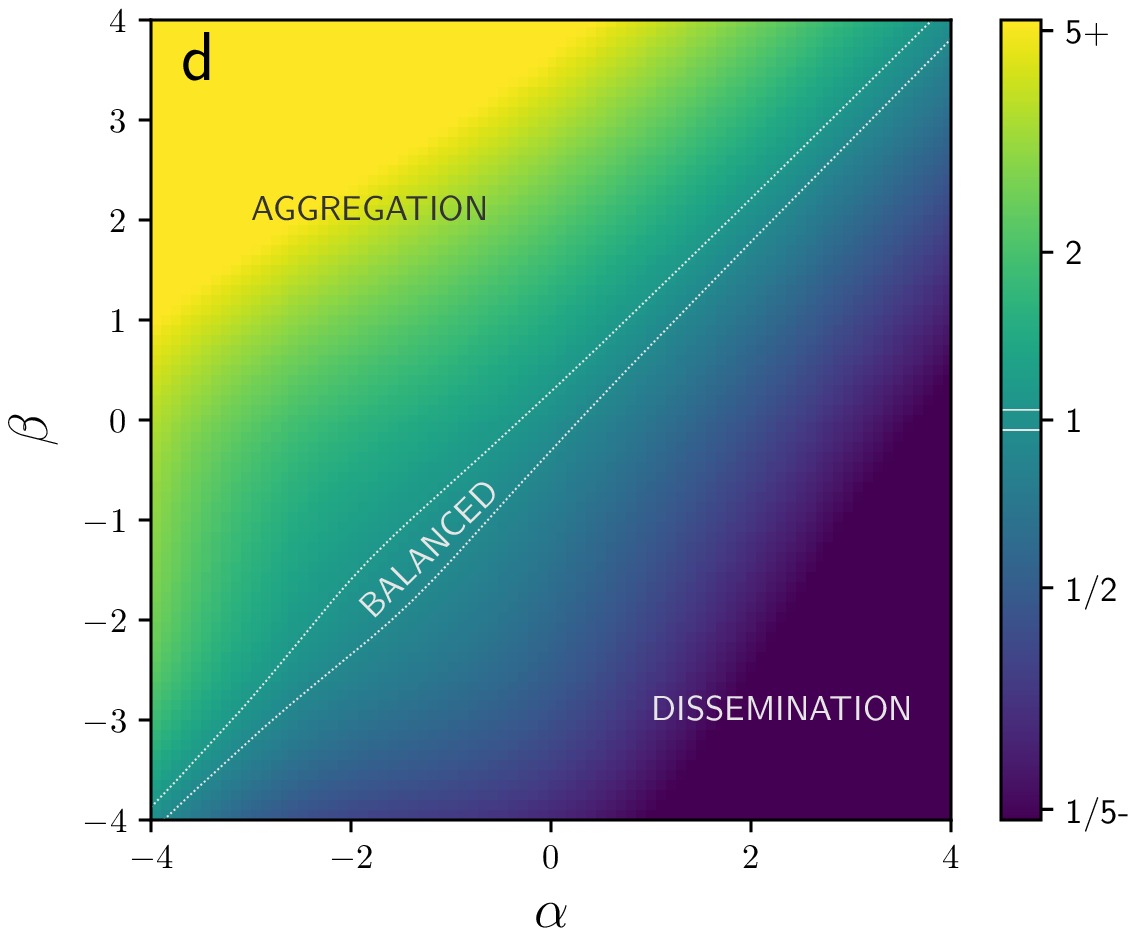}
\caption{Heatmaps of initial slope ratio $S_ {in}/S_{out}$ for \ER networks. Top (a,b): mean degree $25$, bottom (c,d): mean degree $5$.
Left (a,c): exact matrix method results, right (b,d): results using linear approximation.}
\label{fig_ratios_separable}
\end{figure*}

For \ER networks we find a complex dependence of the ratio $S_{in}/S_{out}$ on the exponents $\alpha$ and $\beta$, as can be seen in Figure \ref{fig_ratios_separable}. When $\alpha$ and $\beta$ have similar values, we find balanced infection processes. However, when the two exponents differ significantly, we may find either aggregation (typically when $\beta > \alpha$ and is positive), dissemination ($\beta < \alpha$ and $\alpha$ positive) or again balanced processes. Just as in the case of dependence only on the source or on the destination degree, the exact mapping of the three types of processes depends strongly on the mean degree of the network.
As can be seen in the Figure,  the linear expansion approximation, given in Section \ref{separable} as equation (\ref{ratio_separable}) is very accurate for large mean degrees, but cannot be relied upon for small mean degrees.



\section{Effect of broad degree distributions}\label{SF}

It is well established that highly heterogeneous degree distributions, such as heavy-tailed distributions, can significantly alter the behavior of processes occuring on networks, and may also alter the critical behavior. To explore such effects, we considered the process on random networks with powerlaw distributed degrees.

For concreteness, we consider again the separable form of the transmission rate function $\lambda f(k,k') = \lambda k^{\alpha}k'^{\beta}$ -- i.e. we use equation (\ref{fkk_powers}) with the edge occupation probability $T_{k,k'}$ given by equation (\ref{Tkk}) -- and now use it on random graphs with powerlaw degree distributions: $P(q) \propto q^{-\gamma}$.

\begin{figure}[htpb!]
\centering
\includegraphics[width=0.4\columnwidth]{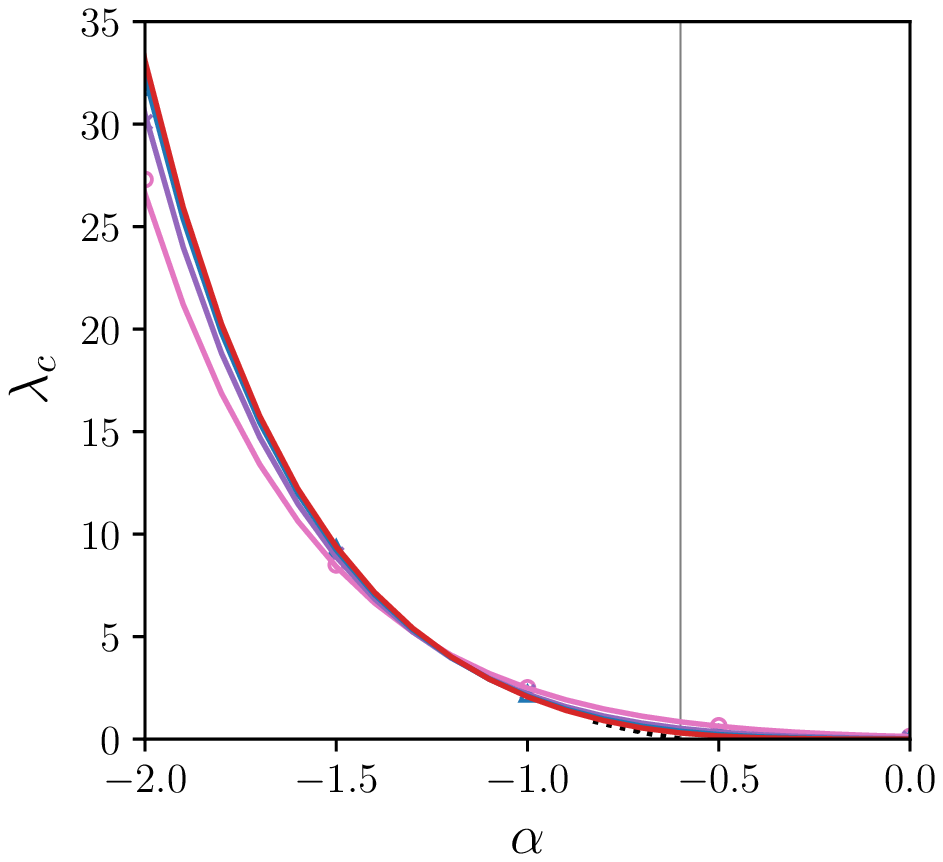}
\includegraphics[width=0.4\columnwidth]{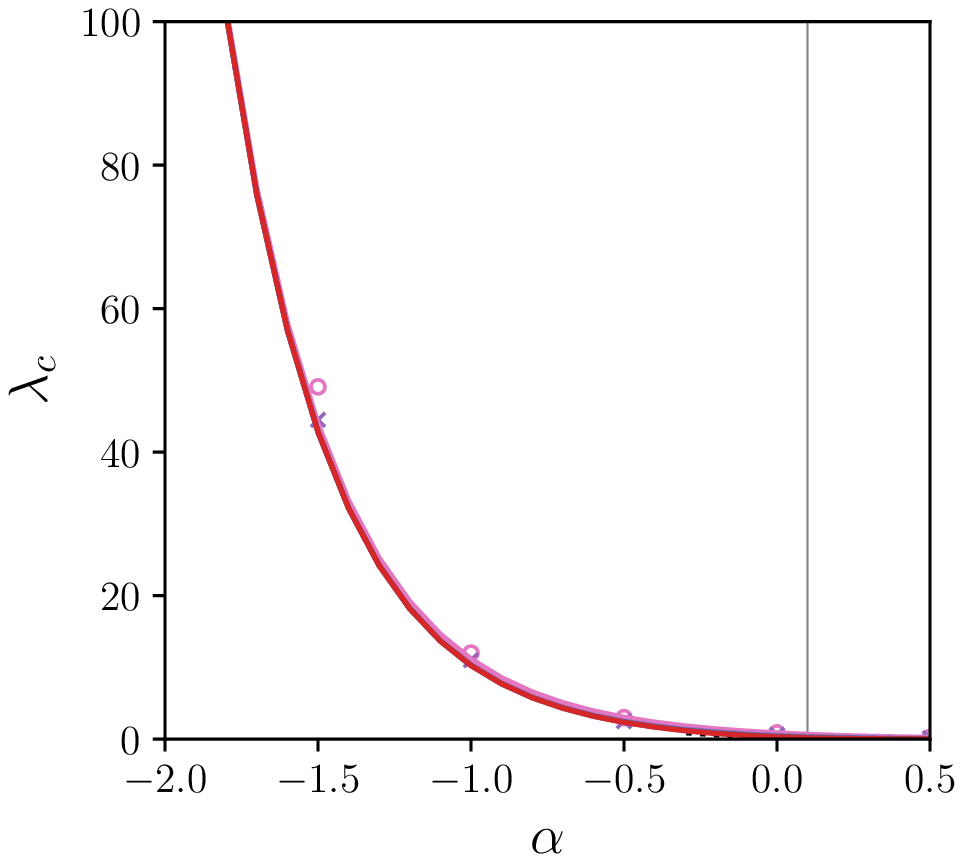}
\includegraphics[width=0.4\columnwidth]{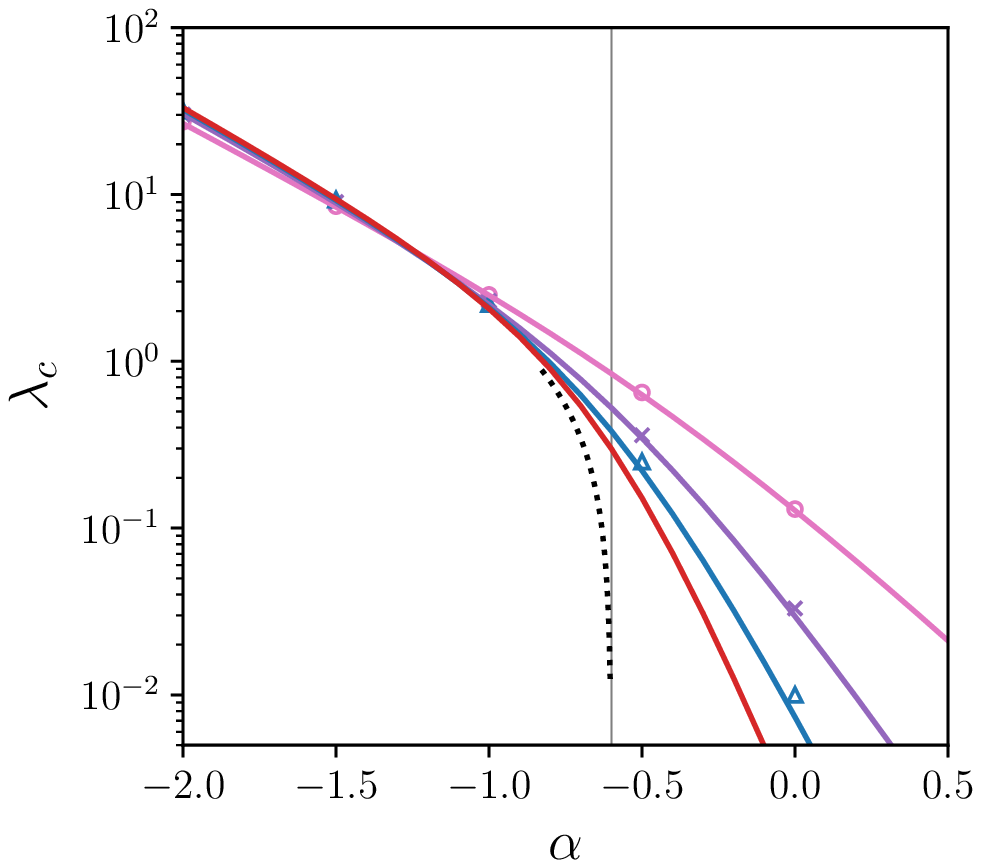}
\includegraphics[width=0.4\columnwidth]{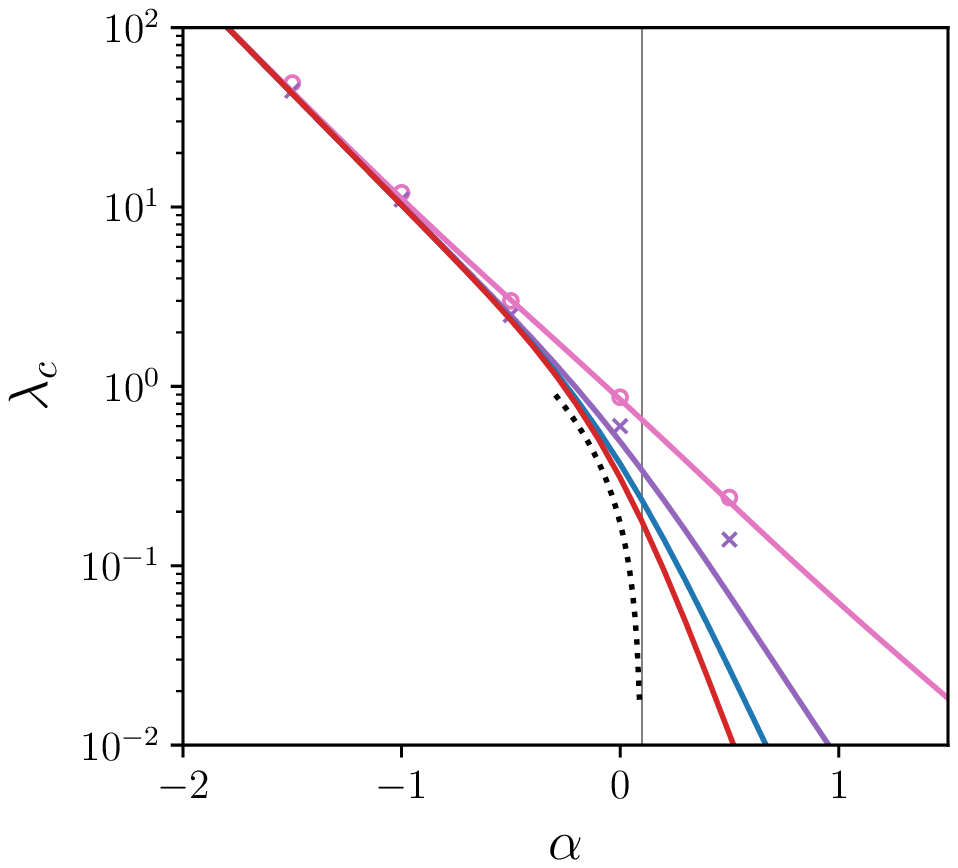}
\caption{Epidemic threshold $\lambda_c$ as a function of $\alpha$ when $\beta=0$.
Upper: linear scale, lower: logarithmic vertical scale.
Dashed black line: analytic linear approximation given by equation (\ref{lamc_separable}).
Solid lines: matrix method calculations for maximum degree cutoffs $k_{max}=100$, $1000$, $10^4$ and $10^5$ from right to left, respectively.
Symbols show simulation results on networks of $N= 10^7$ nodes, with $k_{max} = 100$ (circles) $1000$ (crosses) and $10^4$ (triangles, left panels only).
Left: $\gamma = 2.4$, right $\gamma= 3.1$.
Vertical line marks where $\gamma = 3+\alpha$.
}\label{fig_lamc_SFonlyout}
\end{figure}
One may solve Eqs. (\ref{a_k_matrix}) and (\ref{matrixM}) with a fixed maximum degree $k_{max}$ to obtain the critical threshold $\lambda_c$ using our matrix formulation (see Appendix \ref{MatrixMethod}). 
One must solve a linear equation involving a matrix whose rows and columns correspond to different possible degrees. This method is therefore limited computationally in the maximum degree that can be considered. 
We plot the threshold as a function of $\alpha$ for the case $\beta=0$ in Figure \ref{fig_lamc_SFonlyout}. Results converge quickly with increasing $k_{max}$ in parameter ranges where $\lambda_c$ is not small, indicating that this method is effective in these conditions. We also carried out simulations of the process on networks with $N=10^7$ nodes and powerlaw degree distributions, imposing a degree cutoff to match each value of $k_{max}$ plotted in the figure. 
Note that such finite networks have a natural degree cutoff that depends on the exponent $\gamma$. This means that for the network size we are able to simulate, some larger values of $k_{max}$ are inaccessible. For $\gamma = 2.4$ we obtain reliable results up to $k_max = 10^4$, while for $\gamma = 3.1$, results for $k_{max} = 10^3$ already start to be affected by this limitation.
As expected, we find excellent agreement with the numerical solutions in the cases not affected by the natural degree cutoff.

\begin{figure}[htpb!]
\centering
\includegraphics[width=0.4\columnwidth]{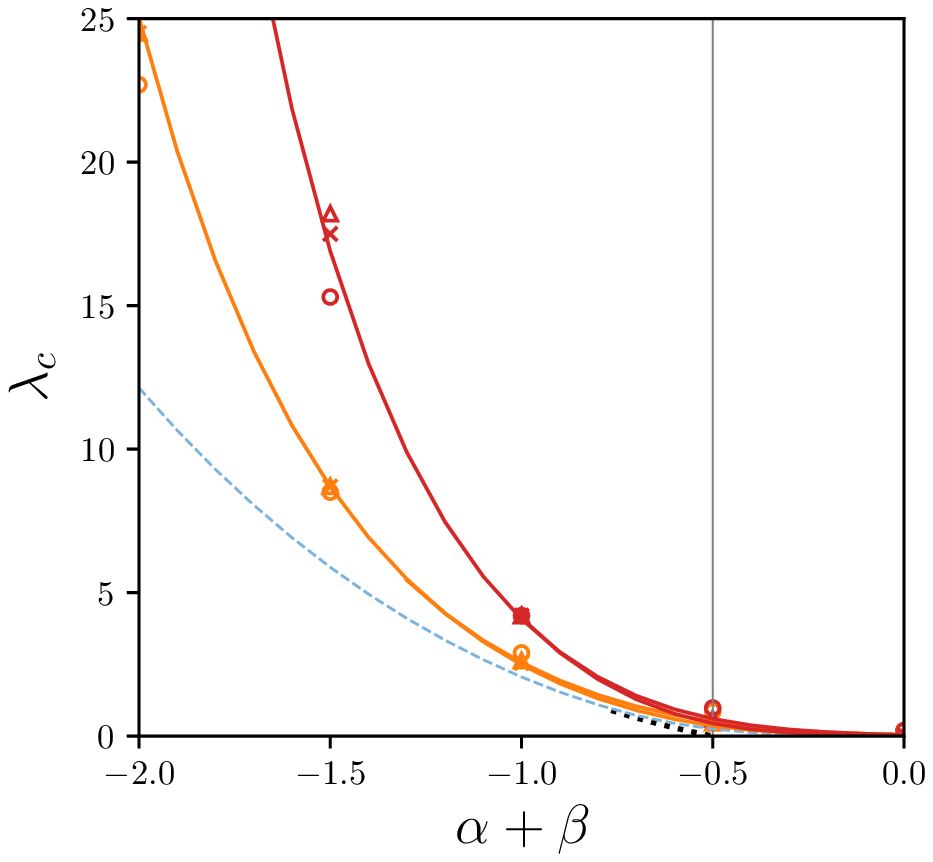}
\includegraphics[width=0.4\columnwidth]{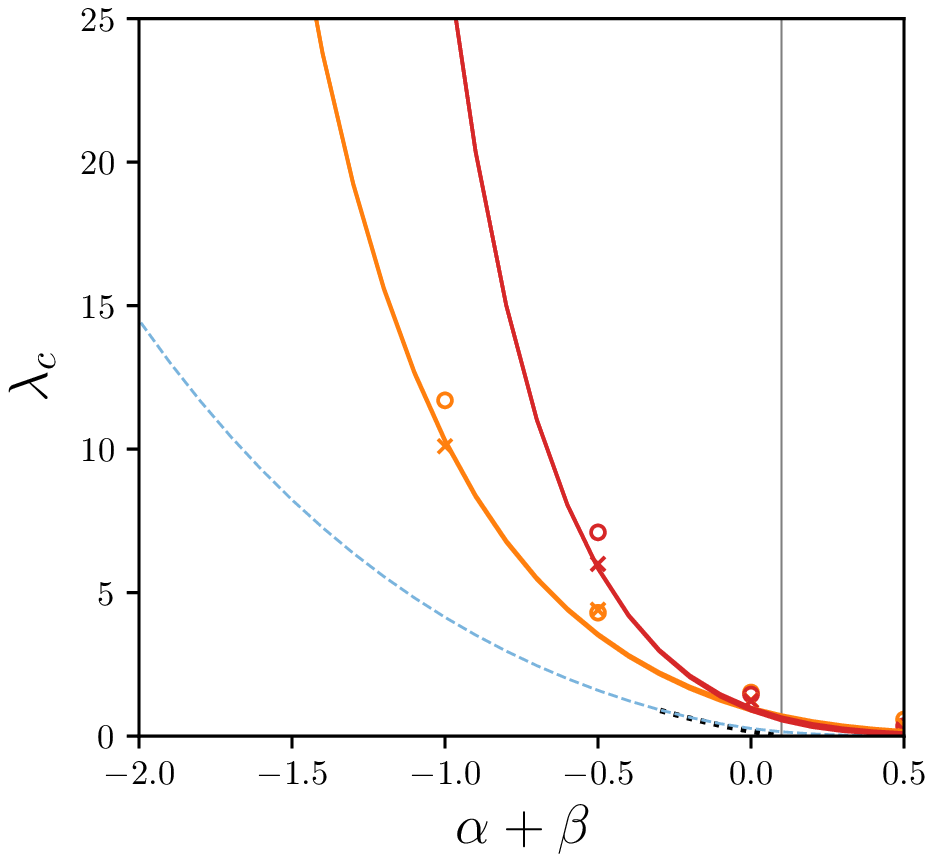}
\includegraphics[width=0.4\columnwidth]{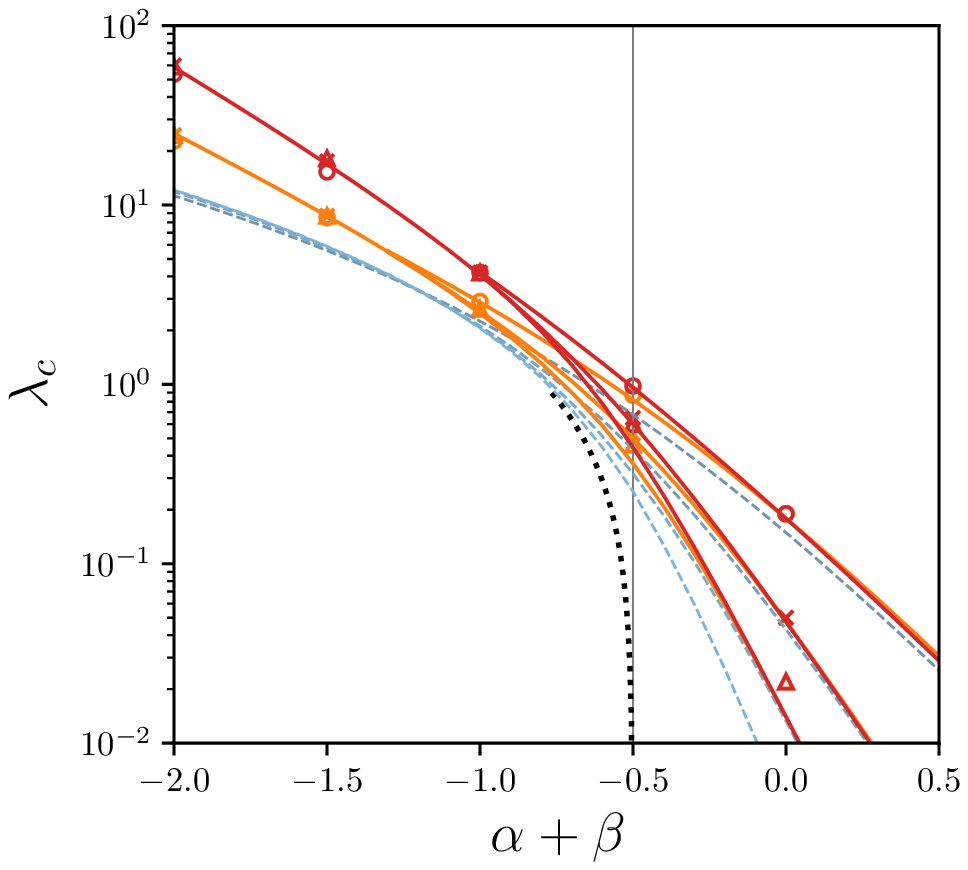}
\includegraphics[width=0.4\columnwidth]{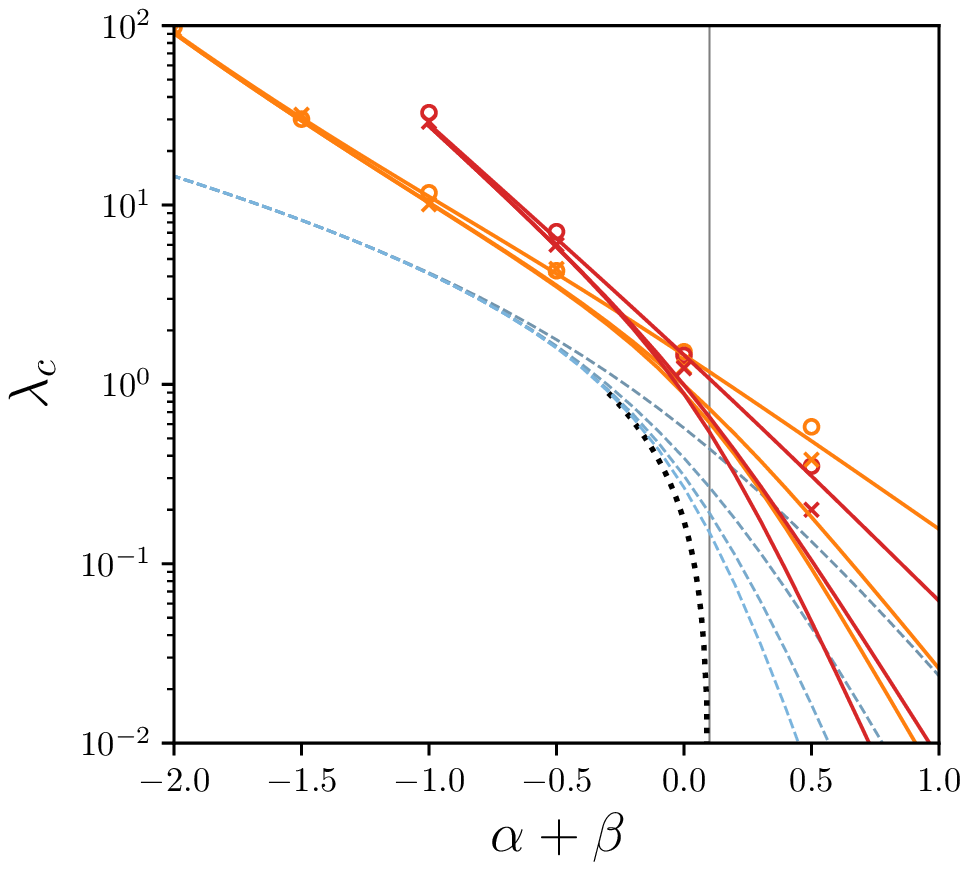}
\caption{
Epidemic threshold $\lambda_c$ as a function of $\alpha+\beta$ (when $\beta\neq 0$).
Upper: linear scale, lower: logarithmic vertical scale.
Solid lines show matrix method results for $\beta = 1$ (red) and $\beta = -1$ (orange), for maximum degrees $100$, $1000$ and $10^4$. 
Symbols show corresponding simulation results on networks of $N= 10^7$ nodes, with $k_{max} = 100$ (circles) $1000$ (crosses) and $10^4$ (triangles, left panels only).
Close to and above the limit $\alpha + \beta = \gamma-3$ (marked by a vertical line) finite size effects are clear on a logarithmic scale.
Dashed blue lines: numerical calculation of linear approximation for maximum degrees, $100$, $1000$, $10^4$ and $10^5$  from right to left, respectively. The dotted black line shows the analytic linear approximation limit for infinite maximum degree, given by equation (\ref{lamc_separable}).
Left: $\gamma = 2.5$, right: $\gamma= 3.1$.
}
\label{fig_lamc_SFboth}
\end{figure}

In Figure \ref{fig_lamc_SFboth} we plot the calculated threshold as a function of $\alpha+\beta$ for two different values of $\beta$. Again, results converge quickly with $k_{max}$, and there is excellent agreement with simulation results, for those cases which are accessible with finite size simulations. We see that, unlike for \ER networks, there is a strong dependence on both $\alpha$ and $\beta$.

As can be seen in the figures, the critical threshold tends to zero at $\gamma = 3+\alpha+\beta$. 
In this regime, it is not practical to obtain exact matrix method solutions for very large values of the maximum degree $k_{max}$. In this case, we may use the linear approximation of Section \ref{separable}, which is valid in precisely this regime. It predicts that the threshold depends only on the sum $\alpha + \beta$. As can be seen in Figure \ref{fig_lamc_SFboth}, matrix method calculations tend to converge, and to agree with numerical solution of the linear approximation equations with a finite maximum degree imposed.

Substituting equation (\ref{fkk_powers}) into equation (\ref{lamc_separable}) gives the epidemic threshold
\begin{align}\label{lamc_powers}
\lambda_c = \frac{\langle k\rangle}{\langle k(k-1) k^{\alpha+\beta}\rangle}\,.
\end{align}
The largest moment in the denominator, $\langle k^{2+\alpha+\beta}\rangle$ is divergent when $\alpha+\beta > \gamma-3$, indicating that the epidemic threshold vanishes at $\gamma = 3+ \alpha + \beta$, confirming a prediction made in \cite{chu2011epidemic}.

Near this point, the threshold is proportional to the distance from  it:
\begin{align}\label{lamc_limit_POW}
\lambda_c \propto {\gamma-3-\alpha-\beta}.
\end{align}
Unlike in, for instance, the percolation transition, the point at which the threshold vanishes depends not only on the structure of the network but also on the degree dependence parameters $\alpha$ and $\beta$.
In fact this degree weighting may effectively counteract the effect of network structure. When $\alpha+\beta +2 = 0$ the results are nearly independent of the degree distribution. This was previously observed for inverse degree dependence $\alpha= \beta = -1$ \cite{olinky2004unexpected}.

\medskip

One may also obtain analytic expressions for the initial growth of the giant-in and -out components above the epidemic threshold, in the region where the threshold is zero, or very close to it, using again the 
linear approximation for the transmission probabilities. As occurs in percolation on powerlaw degree-distributed networks \cite{cohen2002percolation}, this initial growth becomes nonlinear ($\beta$-exponent not equal to one).

For the behaviour of $S_{in}$, 
we expand the right hand side of equation (\ref{A_separable}) for small $A$ (assuming constant infective lifetimes $R(\tau) = \delta(\tau-1)$), in order to find the relationship between $\lambda$ and $A$ in the limit $A \to 0^+$. 
Note that this expansion contains the moment $\langle k^{1+\beta}\rangle$ which diverges for $\gamma < 2+\beta$, hence we are not able to consider this small region (in the case that $\beta > 0$). Furthermore, we consider only $\alpha > -1$.
The expansion has a nonlinear term, which may be the leading order or next-to leading order, depending on the values of $\gamma$, $\alpha$ and $\beta$. 
Having found the behaviour of $A$, we can find the behavior of $S_{in}$ near the transition by making a similar expansion of equation (\ref{Sin_separable}).
More details of these calculations are given in Appendix \ref{PowerlawSlopes}.

We find several regions of the $\gamma\mbox{--}\alpha\mbox{--}\beta$ parameter space with different growth exponents. We may summarise them as follows:

When $\gamma < 3+\alpha+\beta$ the epidemic threshold is $\lambda_c = 0$.  Here we are considering only $\gamma > 2$, and as already stated, the linear approximation is not valid for $\gamma < 2+\beta$, this the lower limit for $\gamma$ that we can consider is either $2$ or $2+\beta$, whichever is larger.
The probability $A$ has the same growth exponent in this whole interval $\max\{2,2+\beta\} < \gamma < 3+\alpha +\beta$. However $S_{in}$ behaves linearly in $\lambda A$ when $\gamma > 2+\alpha$,  and nonlinearly below, giving
\begin{align}\label{slopes_SF_Sin1}
S_{in} &\sim \lambda^{(1+\alpha)/(3-\gamma+\alpha+\beta)} 
\end{align}
for $ \max\{2,2 +\beta,2+\alpha\} < \gamma < 3+\alpha+\beta$,
and
 \begin{align}\label{slopes_SF_Sin2}
S_{in} &\sim \lambda^{(\gamma-1) /(3-\gamma+\alpha+\beta)} 
\end{align}
for $\max\{2,2 +\beta\} < \gamma < 2+\alpha$, which requires that $\alpha > 0$ and $\beta < \alpha$.

When $\gamma > 3+\alpha + \beta$ the epidemic threshold is non-zero, given by equation (\ref{lamc_powers}). However, just above this limit, we may still use the linear approximation to establish the behavior of $S_{in}$ above the epidemic threshold. That is, we may also calculate the growth of $S_{in}$ for $\gamma \to 3+\alpha+\beta^+$.
Again the linear approximation is valid for $\gamma$ larger than $2$ or $2+\beta$, whichever is higher. If $\beta < \alpha$, we again find different exponents on either side of the line $\gamma = 2+\alpha$.  We find that
 \begin{align}\label{slopes_SF_Sin3}
S_{in} &\sim (\lambda-\lambda_c)^{(1+\alpha)/(\gamma-3-\alpha-\beta)}
\end{align}
for $\gamma > 2+\alpha$ in the limit $\gamma \to 3+\alpha+\beta^+$,
and
 \begin{align}\label{slopes_SF_Sin4}
S_{in} &\sim (\lambda-\lambda_c)^{(\gamma-1)/(\gamma-3-\alpha-\beta)}
\end{align}
for $2 < \gamma < 2+\alpha$, again when $\gamma \to 3+\alpha+\beta^+$ (which only occurs for $\beta < -1$).
Details of these calculations are given in Appendix \ref{PowerlawSlopes}.

To obtain the growth exponents for $S_{out}$, notice that (with identical unit lifetimes $R(\tau) = \delta(\tau-1)$),  Eqs. (\ref{B_separable}) and (\ref{Sout_separable}) have exactly the same form as Eqs. (\ref{A_separable}) and (\ref{Sin_separable}), being identical under the exchange of labels $\alpha$ and $\beta$. In fact this is clear from the symmetry of the problem. We may therefore immediately obtain the exponents for $S_{out}$ from those for $S_{in}$ by simply exchanging $\alpha$ and $\beta$.  Note that, for the growth exponent of $S_{out}$, an analytic result using the linear approximation to $T_{k,k'}$ can only be obtained for $\gamma > 2+\alpha$. Below this limit, the moment $\langle k^{1+\alpha}\rangle$ , which appears in the right hand side of equation (\ref{B_separable}), diverges and no solution is possible. We restrict our considerations to $\beta > -1$.
We then obtain, in the region where the threshold is zero,
\begin{align}\label{slopes_SF_Sout1}
S_{out} &\sim \lambda^{(1+\beta)/(3-\gamma+\alpha+\beta)} 
\end{align}
for $ \max\{2,2 +\alpha,2+\beta\} < \gamma < 3+\alpha+\beta$.
This corresponds to a percolation problem with an `effective degree exponent' $\gamma' = \frac{\gamma-\alpha+2\beta}{1+\beta}$, generalizing the expression given in \cite{giuraniuc2005trading,karsai2006nonequilibrium} for the case $\alpha=\beta \equiv -\mu$.
For $\max\{2,2 +\alpha\} < \gamma < 2+\beta$, which requires that $\beta > 0$ and $\alpha < \beta$, we obtain
 \begin{align}\label{slopes_SF_Sout2}
S_{out} &\sim \lambda^{(\gamma-1) /(3-\gamma+\alpha+\beta)}\,. 
\end{align}

When the threshold is not zero
 \begin{align}\label{slopes_SF_Sout3}
S_{out} &\sim (\lambda-\lambda_c)^{(1+\beta)/(\gamma-3-\alpha-\beta)}
\end{align}
for $\gamma > 2+\beta$ in the limit $\gamma \to 3+\alpha+\beta^+$,
and
 \begin{align}\label{slopes_SF_Sout4}
S_{out} &\sim (\lambda-\lambda_c)^{(\gamma-1)/(\gamma-3-\alpha-\beta)}
\end{align}
for $2 < \gamma < 2+\beta$, again when $\gamma \to 3+\alpha+\beta^+$ (which only occurs for $\alpha < -1$).

In the specific case $\alpha = \beta = 0$, we have the standard SIR process, and the exponents are in agreement with those found for the percolation transition \cite{dorogovtsev2008critical, cohen2002percolation}.


\section{Conclusions}\label{conclusions}

In this paper we have investigated SIR epidemic models on configuration model networks, where transmission rates along edges are taken
to be dependent on the degrees of the transmitting and receiving node. The induced heterogeneity in transmission rates requires
us to use a directed percolation mapping to describe the final outcome of the spreading process. One important consequence is that
in such systems the probability of an epidemic occurring may be different from the expected size of an epidemic. 

For arbitrary degree distribution, infective lifetime distribution and an arbitrary matrix of transmission probabilities we have proposed
a numerical method to determine the critical point at which epidemics occur. In addition we have calculated the probability of an epidemic
and the expected epidemic size close to the critical point. These two quantities allow us to categorize epidemic processes as (i)
aggregation processes, where the probability of an epidemic is large, but the fraction of nodes affected is small, (ii) dissemination
processes, where the probability of an epidemic is small, but a large fraction of nodes is affected, or (iii) balanced processes, where
the two quantities are similar in magnitude. Through various examples we have demonstrated that the category of an epidemic process has a
complicated combined dependence on degree distribution and transmission matrix.
For the special case of separable transmission rates we have obtained analytical expressions for the critical point, the epidemic probability
and epidemic size close to the critical point.

We have also studied this epidemic process on random scale-free networks where the transmission rates are separable power functions of transmitting and receiving node degree.
Assuming a small epidemic threshold we have been able to derive its dependence on the degree distribution exponent and the exponents of
the transmission rate function. 
The percolation threshold vanishes in such networks when the asymptotic powerlaw decay exponent $\gamma$ of the degree distribution is below $3$. However with heterogeneous transmission rates this may no longer be the case. We give a condition in terms of $\gamma$, the transmission rate exponent $\alpha$ and the infection rate exponent $\beta$ for the threshold to be finite.
Furthermore, for certain combinations of $\alpha$ and $\beta$ the dependence on the degree distribution is almost entirely removed.
We have also obtained the order parameter critical exponents in this case. We find that both epidemic probability $S_{in}$ and epidemic size $S_{out}$ grow with nonlinear exponents of the distance above the epidemic threshold, and these exponents depend in a complex way on the exponent $\gamma$, $\alpha$ and $\beta$. Interestingly in this model setting the
order parameter exponents of epidemic probability and epidemic size can be different, in contrast to ordinary directed percolation and
previously studied variations of the SIR model.


\appendix


\section{Numerical Method For Finding the Exact Critical Threshold}\label{MatrixMethod}

For a model with given degree-dependent spreading rates $\lambda f_{k, k'}$ (with a given value of $\lambda$), one may first calculate the transmission probabilities,

\begin{align}\label{rates2probs_b}
T_{k, k'} = 1 - e^{- \lambda \tau f_{k,k'}},
\end{align}

\noindent
then plug these values into equation (\ref{matrixM}) to get matrix $\mathbf{M}$. The LEV (largest eigenvalue) of this matrix determines whether our given spreading process is above, at, or below the critical point for the given value of $\lambda$. To find the critical point $\lambda_c$, one may start from an arbitrary $\lambda^{(0)}$ value, and approximate $\lambda_c$ iteratively, according to the following scheme.

Let us denote the LEV of a matrix $\mathbf{A}$ as $\| \mathbf{A} \|$. Let us assume that in the $t^{th}$ approximation of $\lambda$ we are close to the critical point, i.e., $|\lambda^{(t)} - \lambda_c| \ll 1$ and $|\| \mathbf{M}^{(t)} \| - 1| \ll 1$. Let us approximate (using only linear terms) the change in matrix $\mathbf{M}^{(t)}$ as a result of a small change in $\lambda^{(t)}$,

\begin{align}\label{approx1_b}
\delta \mathbf{M}^{(t)} \cong \delta \lambda^{(t)} \frac{ \partial \mathbf{M}^{(t)} }{ \partial \lambda } \Bigr|_{\lambda = \lambda^{(t)}}.
\end{align}
%
Now let us write a right eigenvector equation for the changed matrix $\mathbf{M}^{(t)} + \delta \mathbf{M}^{(t)}$, the changed LEV and the changed principal right eigenvector,
\begin{equation}\label{eig_eq1}
(\mathbf{M}^{(t)} + \delta \mathbf{M}^{(t)}) (\mathbf{v}^{(t)} + \delta \mathbf{v}^{(t)}) =
 (\| \mathbf{M}^{(t)} \| + \delta \| \mathbf{M}^{(t)} \|) (\mathbf{v}^{(t)} + \delta \mathbf{v}^{(t)}).
\end{equation}

\noindent
Neglecting higher-order terms, this reduces to
\begin{align}\label{eig_eq2}
\delta \mathbf{M}^{(t)} \mathbf{v}^{(t)} + \mathbf{M}^{(t)} \delta \mathbf{v}^{(t)} = \delta \| \mathbf{M}^{(t)} \| \mathbf{v}^{(t)} + \| \mathbf{M}^{(t)} \| \delta \mathbf{v}^{(t)}.
\end{align}
%
Multiplying both sides from the left by $\mathbf{w}^{(t)}$, the principal left eigenvector of $\mathbf{M}^{(t)}$, and rearranging, we get
\begin{align}\label{eig_eq3}
\delta \| \mathbf{M}^{(t)} \| = \frac{ \mathbf{w}^{(t)} \delta \mathbf{M}^{(t)} \mathbf{v}^{(t)} }{ \mathbf{w}^{(t)} \mathbf{v}^{(t)} }.
\end{align}

\noindent
Substituting equation (\ref{approx1_b}) into equation (\ref{eig_eq3}) we have
\begin{align}\label{eig_eq4}
\delta \| \mathbf{M}^{(t)} \| = \delta \lambda^{(t)} \frac{ \mathbf{w}^{(t)}    \frac{ \partial \mathbf{M}^{(t)} }{ \partial \lambda } \Bigr|_{\lambda = \lambda^{(t)}}    \mathbf{v}^{(t)} }{ \mathbf{w}^{(t)} \mathbf{v}^{(t)} }.
\end{align}
%
Writing $\delta \lambda^{(t)} = \lambda^{(t+1)} - \lambda^{(t)}$ and requiring $\| \mathbf{M}^{(t+1)} \| = \| \mathbf{M}^{(t)} \| + \delta \| \mathbf{M}^{(t)} \| = 1$, we have
\begin{align}\label{expression_b}
\lambda^{(t+1)} =  \frac{     (1 - \| \mathbf{M}^{(t)} \|)       \mathbf{w}^{(t)} \mathbf{v}^{(t)} }{ \mathbf{w}^{(t)}    \frac{ \partial \mathbf{M}^{(t)} }{ \partial \lambda } \Bigr|_{\lambda = \lambda^{(t)}}    \mathbf{v}^{(t)} } + \lambda^{(t)}.
\end{align}

The critical point $\lambda_c$ is an asymptotically stable attractive fixed point of the operation (\ref{expression_b}). Iterating equation (\ref{expression_b}), generally only a few steps are required to converge to the critical rate $\lambda_c$.
As a reminder let us write the elements of the two matrices involved,
\begin{align}\label{elements1_b}
(\mathbf{M}^{(t)})_{k,k'} = \frac{k'(k'-1)P(k')}{\langle k \rangle} (1 - e^{- \lambda^{(t)} f_{k,k'}}),
\end{align}
\begin{align}\label{elements2_b}
\left( \frac{ \partial \mathbf{M}^{(t)} }{ \partial \lambda } \Bigr|_{\lambda = \lambda^{(t)}} \right)_{k,k'} = \frac{k'(k'-1)P(k')}{\langle k \rangle} f_{k,k'} e^{- \lambda^{(t)} f_{k,k'}}.
\end{align}

\noindent
For any $\lambda^{(t)}$ these matrices can be evaluated and $\| \mathbf{M}^{(t)} \|$ can be found using the power method for example. Here we assume that the matrices involved have a relatively small rank, i.e., the cutoff degree $k_{max}$. A value of $k_{max} \approx 10000$ is still easily managable, bearing in mind that, in general, these matrices are not sparse.

Note that this method may also be applied in the general case of transmission probabilities given as $T = T(\lambda, k, k')$, as long as $\partial T / \partial \lambda$ can be evaluated.\\


\section{General Calculation of Growth of Giant Components above Critical Threshold}\label{GeneralSlopes}

Once the critical point is found using the method of Appendix \ref{MatrixMethod}, we can numerically find the sizes of the giant components near the critical point. This may be done for an arbitrary transmission probability matrix $T_{k,k'}$ and arbitrary infective lifetimes distribution $R(\tau)$. Here, for the sake of simplicity, we present the method for homogeneous infective lifetimes ($\tau_i = 1$ for all nodes $i$), but the general case can be treated analogously using the equations of Section \ref{INeqs}.

In the case of homogeneous infective lifetimes, we only need a single variable $x_k = x_k(\lambda)$, the probability that only a finite number of individuals become infected via a random edge emanating from an infected node of degree $k$.
The recursive equation that we can write for $x_k$ is:
\begin{equation}\label{x_k2}
x_k(\lambda) = \sum_{k'=1}^{\infty} \frac{k' P(k')}{\langle k \rangle} \left[1 - T_{k,k'}(\lambda) + T_{k,k'}(\lambda) x_{k'}^{k'-1} \right].
\end{equation}
The probability that a single infected node selected uniformly at random gives rise to an epidemic (outbreak infecting a finite fraction of the network) is then
\begin{align}\label{Sin2}
S_{in}(\lambda) = \sum_{k=1}^{\infty} P(k) (1 - x_k(\lambda)^k) \,.
\end{align}
Close to the critical point $x_k$ is close to $1$, so we can expand equation (\ref{x_k2}) in powers of $a_k(\lambda) = 1-x_k(\lambda)$:
\begin{equation}\label{a_k_powers2}
a_k(\lambda) =
\sum_{k'=1}^{\infty} \sum_{n=1}^{k'-1} (-1)^{n-1} \binom{k'-1}{n} \frac{k' P(k')}{\langle k \rangle} T_{k,k'}(\lambda) a_{k'}(\lambda)^n.
\end{equation}
Keeping only the linear term in equation (\ref{a_k_powers2}):
\begin{equation}\label{a_k_linear2}
a_k(\lambda) \approx \sum_{k'=1}^{\infty} \frac{k'(k'-1) P(k')}{\langle k \rangle} T_{k,k'}(\lambda) a_{k'}(\lambda).
\end{equation}
We can write equation (\ref{a_k_linear2}) in matrix form as
\begin{align}\label{a_k_matrix2}
\bf{a}(\lambda) = \bf{M}(\lambda)\bf{a}(\lambda)
\end{align}
where the elements of the matrix $\bf{M}(\lambda)$ are
\begin{align}\label{matrixM2}
M_{k,k'}(\lambda) = \frac{k'(k'-1)P(k')}{\langle k\rangle} T_{k,k'}(\lambda).
\end{align}
Equation (\ref{a_k_matrix2}) is a right eigenvector equation for matrix $\mathbf{M}(\lambda)$, with eigenvalue $1$. By definition $\mathbf{M}(\lambda)$ is a non-negative matrix and all components of $\mathbf{a}(\lambda)$ are positive. Assuming that $\mathbf{M}(\lambda)$ is irreducible, the Perron-Frobenius theorem states that there is only one strictly positive right eigenvector, and this corresponds to the largest eigenvalue. Thus the condition for the critical point of equation (\ref{x_k2}): the largest eigenvalue of matrix $\mathbf{M}(\lambda)$ must be $\nu_1 = 1$.
Near the critical point the solutions of equation (\ref{x_k2}) can be written as $x_k \cong 1 - C_{\nu_1} v_k^{(\nu_1)}$, where $v_k^{(\nu_1)}$ are the components of the principal right eigenvector of matrix $\mathbf{M}$. Therefore, near the critical point $S_{in}$ may be written as
\begin{equation}
S_{in} \approx C_{\nu_1} \sum_{k=1}^{\infty} k P(k) v_k^{(\nu_1)}.
\label{eq:640b}
\end{equation}

To find the dependence of the coefficient $C_{\nu_1}$ on the distance from the critical point we need to consider also quadratic terms on the right hand side of equation (\ref{a_k_powers2}),
\begin{multline}\label{a_k_quadratic}
a_k(\lambda) \approx \sum_{k'=1}^{\infty} \frac{k'(k'-1) P(k')}{\langle k \rangle} T_{k,k'}(\lambda) a_{k'}(\lambda) \\
- \sum_{k'=1}^{\infty} \frac{k'(k'-1) (k'-2) P(k')}{2 \langle k \rangle} T_{k,k'}(\lambda) a_{k'}(\lambda)^2,
\end{multline}

\noindent
which may be written in matrix form as

\begin{equation}
\mathbf{a} = \mathbf{M} \mathbf{a} - \mathbf{Q} \mathbf{a}^{\circ 2},
\label{eq:641b}
\end{equation}

\noindent
where matrix $\mathbf{Q}$ has elements

\begin{align}\label{matrixQ2}
Q_{k,k'} = \frac{k'(k'-1)(k'-2)P(k')}{2 \langle k\rangle} T_{k,k'},
\end{align}

\noindent
and $\mathbf{a}^{\circ 2}$ denotes Hadamard (element-wise) square of vector $\mathbf{a}$.
The set of right eigenvectors of matrix $\mathbf{M}$ forms a complete basis in which any vector (of the correct dimension) can be expressed as a linear combination of these vectors. Let us express $\mathbf{a}$ in such a way,

\begin{equation}
\mathbf{a} = \sum_{\nu} C_{\nu} \mathbf{v}^{(\nu)},
\label{eq:650b}
\end{equation}

\noindent
where $\mathbf{v}^{(\nu)}$ is the right eigenvector of $\mathbf{M}$ corresponding to eigenvalue $\nu$.
Substituting (\ref{eq:650b}) into equation (\ref{eq:640b}), we have

\begin{equation}
\sum_{\nu} C_{\nu} \mathbf{v}^{(\nu)} = \sum_{\nu} C_{\nu} \mathbf{M} \mathbf{v}^{(\nu)} - \sum_{\nu, \nu'} C_{\nu} C_{\nu'} \mathbf{Q} (\mathbf{v}^{(\nu)} \circ \mathbf{v}^{(\nu')}),
\label{eq:660b}
\end{equation}

\noindent
where $\circ$ denotes the Hadamard product.
Since $\mathbf{v}^{(\nu)}$ is an eigenvector of $\mathbf{M}$ with eigenvalue $\nu$, equation (\ref{eq:660b}) simplifies to

\begin{equation}
\sum_{\nu} C_{\nu} (\nu - 1) \mathbf{v}^{(\nu)} = \sum_{\nu, \nu'} C_{\nu} C_{\nu'} \mathbf{Q} (\mathbf{v}^{(\nu)} \circ \mathbf{v}^{(\nu')}).
\label{eq:670b}
\end{equation}

\noindent
Let us remember that the left eigenvectors $\mathbf{w}^{(\nu)}$ and the right eigenvectors $\mathbf{v}^{(\nu)}$ of matrix $\mathbf{M}$ must fulfil $\mathbf{w}^{(\nu)} \mathbf{v}^{(\nu')} = 0$ if $\nu \neq \nu'$. With this in mind, we multiply both sides of equation (\ref{eq:670b}) by $\mathbf{w}^{(\nu_1)}$, and keep only leading order terms on the right hand side,

\begin{equation}
C_{\nu_1} (\nu_1 - 1) \mathbf{w}^{(\nu_1)}  \mathbf{v}^{(\nu_1)} = C_{\nu_1}^2 \mathbf{w}^{(\nu_1)} \mathbf{Q} (\mathbf{v}^{(\nu_1)})^{\circ 2}.
\label{eq:680b}
\end{equation}

\noindent
Thus the coefficient $C_{\nu_1}$ is

\begin{equation}
C_{\nu_1} = \frac{ (\nu_1 - 1) \mathbf{w}^{(\nu_1)} \mathbf{v}^{(\nu_1)} }{ \mathbf{w}^{(\nu_1)} \mathbf{Q} (\mathbf{v}^{(\nu_1)})^{\circ 2} },
\label{eq:690b}
\end{equation}

\noindent
which, as expected, is linear in terms of the distance $\nu_1 - 1$ from the critical point.
We can now express the size of the giant in-component near the critical point,

\begin{align}
S_{in} &= \sum_{k=1}^{\infty} P(k) (1 - x_k^k) \cong C_{\nu_1} \sum_{k=1}^{\infty} k P(k) v_k^{(\nu_1)}\\
&\cong (\nu_1 - 1) \frac{ (\mathbf{d} \mathbf{v})  (\mathbf{w} \mathbf{v}) }{ \mathbf{w} \mathbf{Q} \mathbf{v}^{\circ 2} }, \label{Sin_slope}
\end{align}

\noindent
where we introduced the vector $\mathbf{d}$ with components $d_k = kP(k)$. For the sake of simplicity we have omitted the superscripts $(\nu_1)$ from the principal left and right eigenvectors $\mathbf{w}$ and $\mathbf{v}$. The corresponding expression for the size of the giant out-component is

\begin{equation}\label{Sout_slope}
S_{out} \cong (\nu_1 - 1) \frac{  (\mathbf{d} \mathbf{v'})    (\mathbf{w'} \mathbf{v'}) }{ \mathbf{w'} \mathbf{Q'} \mathbf{v'}^{\circ 2} },
\end{equation}

\noindent
where $\mathbf{w'}$ and $\mathbf{v'}$ are the left and right principal eigenvectors of the matrix $\mathbf{M'}$, given as

\begin{align}\label{matrixM2b}
M'_{k,k'} = \frac{k'(k'-1)P(k')}{\langle k\rangle} T_{k',k},
\end{align}

and the elements of matrix $\mathbf{Q'}$ are

\begin{align}\label{matrixQ2b}
Q'_{k,k'} = \frac{k'(k'-1)(k'-2)P(k')}{2 \langle k\rangle} T_{k',k}.
\end{align}

When the matrix $\mathbf{T}$ of transmission probabilities is symmetric - e.g., in the original SIR model-, Eqs. (\ref{matrixM2}) and (\ref{matrixQ2}) coincide with Eqs. (\ref{matrixM2b}) and (\ref{matrixQ2b}), and the sizes of the giant components are the same.
For arbitrary $\mathbf{T}$ and $P(k)$, Eqs. (\ref{Sin_slope}) and (\ref{Sout_slope}) may be evaluated numerically to determine what category the given spreading process falls into (see Fig \ref{fig:30}).
Note that equation (\ref{a_k_powers2}) may be used to derive closer approximations to the sizes of the giant components in terms of higher powers of $\nu_1 - 1$.

It is more convenient to take the rate parameter $\lambda$ as control parameter, instead of the LEV of the $\mathbf{M}(\lambda)$ matrix. To make the conversion, we simply need to consider equation (\ref{eig_eq4}) that relates an infinitesimal change in $\lambda$ to the corresponding change in the LEV of a matrix $\mathbf{M}$ dependent on $\lambda$.

\begin{equation}
S_{in} \cong (\lambda - \lambda_c) \frac{  ( \mathbf{d} \mathbf{v} )    (   \mathbf{w}  \frac{ \partial \mathbf{M} }{ \partial \lambda } \Bigr|_{\lambda = \lambda_c}   \mathbf{v} )    }{     \mathbf{w} \mathbf{Q} \mathbf{v}^{\circ 2}   },
\label{slope_in_generic}
\end{equation}

\noindent
and

\begin{align}
S_{out} \cong (\lambda - \lambda_c) \frac{  ( \mathbf{d} \mathbf{v'} )    (   \mathbf{w'}  \frac{ \partial \mathbf{M'} }{ \partial \lambda } \Bigr|_{\lambda = \lambda_c}   \mathbf{v'} )    }{     \mathbf{w'} \mathbf{Q'} \mathbf{v'}^{\circ 2}   }.
\label{slope_out_generic}
\end{align}

The mean relative size of an outbreak initiated at a single node goes as
\begin{equation}
S = S_{in} S_{out} \propto (\lambda - \lambda_c)^2.
\label{slope_all}
\end{equation}


\section{Critical Threshold and Initial Growth rates for powerlaw distributions}

To consider the critical behavior for broad degree distributions, we consider powerlaw degree distributions $P(q) \sim Dk^{-\gamma}$, for $k>k_0$. The moment $\langle k^m\rangle$ diverges in the infinite size limit if $\gamma < m+1$.

We consider
transmission rates $T_{k,k'} = 1 - e^{-\lambda k^{\alpha}k'^{\beta}}$. All lifetimes are assumed to be equal, $\tau=1$ for simplicity.
For $\lambda$ sufficiently small, let us assume that it is reasonable to linearise $T_{k,k'}$, writing
 \begin{equation}\label{Tkk_linearised}
 T_{k,k'} \approx \lambda k^{\alpha}k'^{\beta}.
 \end{equation}
We only consider $\alpha > -1$. Exploration of larger negative values is left for future work.

 \label{PowerlawCriticalPoint}

 The equations for the giant -in component then become
 \begin{equation}
 S_{in} = 1 - \sum_k P(k) [1 - \lambda \frac{\langle k^{1+\beta}\rangle}{\langle k\rangle}k^{\alpha}A]^{k}
 \label{Sin_both}
 \end{equation}
 with
 \begin{equation}\label{A_both}
 A = 1 - \sum_k \frac{k^{1+\beta}P(k)}{\langle k^{1+\beta}\rangle} [1 - \lambda \frac{\langle k^{1+\beta}\rangle}{\langle k\rangle}k^{\alpha}A]^{k-1}.
 \end{equation}
These equations only make sense if the moment $\langle k^{1+\beta}\rangle$ is finite,  which imposes the condition  $\gamma > 2+\beta$.

 Linearising, we obtain the epidemic threshold:
 \begin{equation}
 \lambda_c = \frac{\langle k\rangle}{\langle k^{2+\alpha+\beta} \rangle -  \langle k^{1+\alpha+\beta} \rangle}
 \label{lamcBOTH}
 \end{equation}
the leading moment in the denominator, $\langle k^{2+\alpha+\beta} \rangle $, diverges for $\gamma < 3+\alpha +\beta$ indicating the threshold goes to zero at this point.

\label{PowerlawSlopes}


Returning to equation (\ref{A_both}), we substitute the powerlaw form of the degree distribution and expand to higher order to find the behaviour of $A$ and hence $S_{in}$ near the threshold.
Writing 
$[1 - \lambda B k^{\alpha}A]^{k-1} \approx e^{- \lambda AB k^{1+\alpha}}$
, and approximating the sum over $k$ by an integral, we have
\begin{align}
A 
& \approx
1 
- 
\frac{D}{B\langle k\rangle}
\int_{k_0}^{\infty} k^{1+\beta-\gamma} e^{- \lambda   AB k^{1+\alpha}}dk\,,
\end{align}
where for compactness we have written $B \equiv \langle k^{1+\beta }\rangle / \langle k \rangle$.

We make the substitution $y = B \lambda Ak^{\alpha+1}$,
obtaining
\begin{equation}
A \approx
1 - 
\frac{D}{\langle k \rangle B (\alpha+1)}
\left(\lambda A B\right)^{(\gamma-2-\beta)/(1+\alpha)}
\int_{\lambda AB k_0^{1+\alpha}}^{\infty}  y^{(1-\gamma-\alpha+\beta)/(1+\alpha)} e^{-y} dy\,.
\label{A_integralBOTH}
\end{equation}
We then integrate  (\ref{A_integralBOTH}) by parts one or more times to obtain the leading-order terms.

\medskip

For $2+\beta < \gamma < 3+\alpha+\beta$, the threshold is zero and the leading term in the expansion is the singular term
\begin{align}
A &\cong C (\lambda A)^{(\gamma-2-\beta)/(1+\alpha)}\,, \label{3cBOTH}
\end{align}
where $C=\frac{D(1+\alpha)}{B\langle k\rangle (\gamma-2-\beta)} \Gamma\left(\frac{3-\gamma+\alpha+\beta}{1+\alpha} \right) B^{(\gamma-2-\beta)/(1+\alpha)}$.

\medskip

For $3+\alpha+\beta < \gamma < 4+2\alpha+\beta$ we have a linear then the nonlinear term, whose exponent will be between $1$ and $2$:
\begin{align}
A &\cong
 \frac{\langle k^{2+\alpha+\beta}\rangle}{\langle k \rangle}\lambda A
+ 
C (\lambda A)^{(\gamma-2-\beta)/(1+\alpha)}\,. \label{1cBOTH}
\end{align}
Notice that the coefficient of the linear term diverges if $\gamma < 3+\alpha+\beta$, which is the point at which the nonlinear term becomes of leading order.
Although this form for the expansion is correct across this whole range, our linear approximation is only valid close to the lower limit, where the threshold is arbitrarily close to zero.

Now let us consider the behaviour of $A$ above the threshold.
For $3+\alpha+\beta < \gamma < 4+2\alpha+\beta$, equation (\ref{1cBOTH}) applies, and we have a nonzero threshold $\lambda_c$.
Writing $\lambda = \lambda_c + \epsilon$ in (\ref{1cBOTH}), and using (\ref{lamcBOTH}) gives
\begin{equation}\label{1c_2BOTH}
\frac{1}{\lambda_c}\epsilon A  \approx -C [(\lambda_c+\epsilon)A]^{(\gamma-2-\beta)/(1+\alpha)}.
\end{equation}
Note that, in this region, $C$ is negative.
Now, for very small $\epsilon$, we can write $(\lambda_c + \epsilon)^{\rho} \approx \lambda_c^{\rho} +\epsilon\rho\lambda_c^{\rho-1}$.
Rearranging for $A$ and neglecting higher order terms we find
\begin{equation}\label{1c_5BOTH}
A \cong 
 (-C)^{-(1+\alpha)/(\gamma-3-\alpha-\beta)} \lambda_c^{-(\gamma-1+\alpha-\beta)/(\gamma-3-\alpha-\beta)} 
 \epsilon^{(1+\alpha)/(\gamma-3-\alpha-\beta)}\,.
\end{equation}

\medskip

When $2 +\beta < \gamma < 3+\alpha+\beta$, equation (\ref{3cBOTH}) applies. 
Note that $C$ is positive in this range of $\gamma$. Rearranging equation (\ref{3cBOTH}) to find $A$ in terms of $\lambda$ we find
\begin{align}\label{1c_5BOTH2}
A \cong C^{(1+\alpha)/(3-\gamma+\alpha+\beta)}\lambda^{(\gamma-2-\beta)/(3-\gamma+\alpha+\beta)}\,.
\end{align}

\medskip

Now we need to establish the relationship between the growth exponent for $A$ and the one for $S_{in}$.
We return to equation (\ref{Sin_both}) and make the same approximations and substitutions as before, 
 \begin{equation}
 S_{in} 
\approx 1 - \frac{D}{1+\alpha} (\lambda B A)^{(\gamma-1)/(1+\alpha)}
 \int_{\lambda A Bq_0^{1+\alpha}}^{\infty} y^{(-\gamma-\alpha)/(1+\alpha)} e^{-y} dy\,
 \end{equation}
 with $y = B \lambda Ak^{\alpha+1}$.
Integrating by parts allows us to find the leading order behavior with respect to $\lambda A$.

For $\gamma > 2+\alpha$ we find a linear leading order term
\begin{align}
S_{in} &\cong 
 \langle k^{1+\alpha}\rangle \lambda A B \label{SIn_vs_A1}
\end{align}
while for $\gamma < 2+\alpha$ the nonlinear term is leading
\begin{align}
S_{in} &\cong \hat{C} (\lambda AB)^{(\gamma-1)/(1+\alpha)}\, \label{SIn_vs_A2}
\end{align}
with $\hat{C} = \frac{D}{\gamma-1} \Gamma\left(\frac{2-\gamma+\alpha}{1+\alpha}\right)$.

Combining Eqs. (\ref{SIn_vs_A1}) and (\ref{SIn_vs_A2}) with (\ref{1c_5BOTH}) and (\ref{1c_5BOTH2}), we obtain the growth of $S_{in}$ above the epidemic threshold in four regions, as given in Eqs. (\ref{slopes_SF_Sin1})--(\ref{slopes_SF_Sin4}).

 \bigskip


Using once again the form equation (\ref{Tkk_linearised}), i.e. $g_kh_{k'} = k^{\alpha}k'^{\beta}$ 
The size of the giant -out component is given by
\begin{align}\label{Sout_separableSF}
S_{out} = 1 - \sum_{k} P(k)
\left[1 - \lambda k^{\beta} \frac{\langle k^{1+\alpha}\rangle}{\langle k \rangle}B \right]^{k}\,.
\end{align}
where
\begin{align}\label{B_separableSF2}
B =  1 -\sum_{k} \frac{k^{1+\alpha}P(k)}{\langle k^{1+\alpha}\rangle} 
\left[1 - \lambda  \frac{\langle k^{1+\alpha}\rangle}{\langle k \rangle}k^{\beta} B \right]^{k-1}\,.
\end{align}
Comparing with equation (\ref{A_both})
we see that the equation for $B$ is identical to that for $A$, under the exchange of $\alpha$ and $\beta$.
Futhermore, equation (\ref{Sout_separableSF}) is identical to equation (\ref{Sin_both}) under the same exchange of parameters.
The results Eqs. (\ref{slopes_SF_Sout1})--(\ref{slopes_SF_Sout4}) then immediately follow.

\bigskip


\bibliography{bibliography_SIR}

\providecommand{\newblock}{}
\begin{thebibliography}{10}
\expandafter\ifx\csname url\endcsname\relax
  \def\url#1{{\tt #1}}\fi
\expandafter\ifx\csname urlprefix\endcsname\relax\def\urlprefix{URL }\fi
\providecommand{\eprint}[2][]{\url{#2}}

\bibitem{anderson1992infectious}
Anderson R~M, Anderson B and May R~M 1992 {\em Infectious diseases of humans:
  dynamics and control\/} (Oxford university press)

\bibitem{pastor-satoras2001epidemic}
Pastor-Satorras R and Vespignani A 2001 {\em Physical review letters\/} {\bf
  86} 3200

\bibitem{moreno2002epidemic}
Moreno Y, Pastor-Satorras R and Vespignani A 2002 {\em The European Physical
  Journal B-Condensed Matter and Complex Systems\/} {\bf 26} 521--529

\bibitem{newman2002spread}
Newman M~E~J 2002 {\em Physical Review E\/} {\bf 66} 016128

\bibitem{keeling2005networks}
Keeling M~J and Eames K~T 2005 {\em Journal of the Royal Society Interface\/}
  {\bf 2} 295--307

\bibitem{pastor2015epidemic}
Pastor-Satorras R, Castellano C, Van~Mieghem P and Vespignani A 2015 {\em
  Reviews of modern physics\/} {\bf 87} 925

\bibitem{boguna2002epidemic}
Bogun{\'a} M and Pastor-Satorras R 2002 {\em Physical Review E\/} {\bf 66}
  047104

\bibitem{grassberger1983critical}
Grassberger P 1983 {\em Mathematical Biosciences\/} {\bf 63} 157--172

\bibitem{kenah2007network}
Kenah E and Robins J~M 2007 {\em Journal of theoretical biology\/} {\bf 249}
  706--722

\bibitem{miller2007epidemic}
Miller J~C 2007 {\em Physical Review E\/} {\bf 76} 010101

\bibitem{rogers2015assessing}
Rogers T 2015 {\em EPL (Europhysics Letters)\/} {\bf 109} 28005

\bibitem{stein2011super}
Stein R~A 2011 {\em International Journal of Infectious Diseases\/} {\bf 15}
  e510--e513

\bibitem{moslonka2012weighting}
Moslonka-Lefebvre M, Bonhoeffer S and Alizon S 2012 {\em Journal of theoretical
  biology\/} {\bf 311} 46--53

\bibitem{karsai2006nonequilibrium}
Karsai M, Juh{\'a}sz R and Igl{\'o}i F 2006 {\em Physical Review E\/} {\bf 73}
  036116

\bibitem{britton2011weighted}
Britton T, Deijfen M and Liljeros F 2011 {\em Journal of statistical physics\/}
  {\bf 145} 1368--1384

\bibitem{ferreira2016collective}
Ferreira S~C, Sander R~S and Pastor-Satorras R 2016 {\em Physical Review E\/}
  {\bf 93} 032314

\bibitem{giuraniuc2005trading}
Giuraniuc C, Hatchett J, Indekeu J, Leone M, Castillo I~P, Van~Schaeybroeck B
  and Vanderzande C 2005 {\em Physical review letters\/} {\bf 95} 098701

\bibitem{joo2004behavior}
Joo J and Lebowitz J~L 2004 {\em Physical Review E\/} {\bf 69} 066105

\bibitem{olinky2004unexpected}
Olinky R and Stone L 2004 {\em Physical Review E\/} {\bf 70} 030902

\bibitem{chu2011epidemic}
Chu X, Zhang Z, Guan J and Zhou S 2011 {\em Physica A: Statistical Mechanics
  and its Applications\/} {\bf 390} 471--481

\bibitem{castellano2006non}
Castellano C and Pastor-Satorras R 2006 {\em Physical review letters\/} {\bf
  96} 038701

\bibitem{singh2013nonlinear}
Singh A and Singh Y 2013 {\em Acta Physica Polonica B\/} {\bf 44} 5

\bibitem{cohen2002percolation}
Cohen R, Ben-Avraham D and Havlin S 2002 {\em Physical Review E\/} {\bf 66}
  036113

\bibitem{dorogovtsev2008critical}
Dorogovtsev S~N, Goltsev A~V and Mendes J~F~F 2008 {\em Rev. Mod. Phys.\/} {\bf
  80} 1275

\end{thebibliography}
\bibliographystyle{iopart-num}

\end{document}